\documentclass[aps,prd,eqsecnum,fleqn,preprint]{revtex4}
\usepackage{amsmath,amssymb,amsfonts,verbatim,ifthen}
\usepackage[dvips]{graphicx}
\usepackage{color}

\newcommand\nc{\newcommand}
\nc\nn{\notag}
\nc\cit[1]{Ref.~\cite{#1}}
\nc\cits[1]{Refs.~\cite{#1}}
\nc\refeqn[1]{eqn.~\eqref{#1}}
\nc\rf[1]{(\ref{#1})}
\renewcommand\cal[1]{\mathcal{#1}}
\nc\spa[2]{\langle#1\, #2\rangle}
\nc\spb[2]{[#1\, #2]}
\nc\lp[2]{#1 \cdot #2}
\nc\spRa[2]{\langle#1\, #2]}
\nc\ang[1]{|#1\rangle}
\nc\bra[1]{|#1]}
\nc\bang[1]{\langle#1|}
\nc\bbra[1]{[#1|}
\nc\ab[3]{\langle#1|#2|#3]}
\nc\sandpmmp[3]{%
\left\langle#1^{\pm}\right|{#2}%
\left|#3^{\mp}\right\rangle}
\nc\slashvar[1]{\hbox{$\slash \hskip -.19cm #1$}}
\nc\s[1]{\slashvar#1}%
\nc\xout[1]{\hbox{$\times \hskip -.19cm #1$}}
\nc\pslash{\hbox{$\slash \hskip -.22cm p$}}
\nc\Pslash{\hbox{$\slash \hskip -.24cm P$}}
\nc\epslash{\hbox{$\slash \hskip -.20cm \epsilon$}}
\nc\wh{\widehat}
\nc\Phat{\wh P}
\nc\mhvbartext{\overline{\textrm{MHV}}}
\nc\mhvbarthree{\overline{\textrm{MHV}}_3}
\nc\nxmhv[1]{\textrm{N}^{#1}\textrm{MHV}}
\nc\nnmhv{\textrm{N}^2\textrm{MHV}}
\nc\mhvt{\textrm{MHV}}
\nc\nmhv{\textrm{NMHV}}
\nc\deltfour[1]{\delta^{(4)}(#1)}
\nc\delteight[1]{\delta^{(8)}(#1)}
\nc\deltsixteen[1]{\delta^{(16)}(#1)}
\nc\delteightfactor[4]{\spa{#1}{#2}^4
	\deltfour{\eta_{#1}
		-\sum_{#3}^{#4}\frac{\spa{#2}{i} }{\spa{#1}{#2} }
		\eta_{i} }
	\deltfour{\eta_{#2}
		-\sum_{#3}^{#4}\frac{\spa{#1}{i} }{\spa{#1}{#2} }
		\eta_{i} }  }
\nc\delteightfactorsum[5]{\spa{#1}{#2}^4
	\deltfour{\eta_{#1}
		-\sum_{#5=#3}^{#4}\frac{\spa{#2}{#5} }{\spa{#1}{#2} }
		\eta_{#5} } 
	\deltfour{\eta_{#2}
		-\sum_{#5=#3}^{#4}\frac{\spa{#1}{#5} }{\spa{#1}{#2} }
		\eta_{#5} }  }
\nc\deltmhvbar[3]{\eta_{#1}\spb{#2}{#3}+\eta_{#2}\spb{#3}{#1}
		 +\eta_{#3}\spb{#1}{#2}}
\nc\mhvbar[3]{\frac{\deltfour{\deltmhvbar{#1}{#2}{#3}} }
	{\spb{#1}{#2} \spb{#2}{#3} \spb{#3}{#1}} }
\nc\mhvbarsugra[3]{\frac{\delteight{\deltmhvbar{#1}{#2}{#3}} }
	{\left(\spb{#1}{#2} \spb{#2}{#3} \spb{#3}{#1}\right)^2 } }
\nc\mhv[2]{\frac{\delteight{#1}}{#2}}
\nc\mhvn[4]{\frac{
	\delteight{\susyq{#1}-\susyq{#2}+\sum_{#3}^{#4}\susyq{i}} }
	{\mhvdenom{#2}{#1}{#3}{#4} } }
\nc\mhvnum[4]{\delteight{-\susyq{#1}+\susyq{#2}+\sum_{#3}^{#4}\susyq{i}} }
\nc\mhvnumsum[5]{\delteight{\susyq{#1}-\susyq{#2}+\sum_{#5=#3}^{#4}\susyq{#5}} }
\nc\mhvdenom[4]{\spa{#1}{#2}\spa{#2}{#3}\ldots\spa{#4}{#1} }
\nc\mhvb[2]{\frac{\deltfour{#1}}{#2}}
\nc\mhvsugra[4]{\frac{ 
	\deltsixteen{\susyq{#1}-\susyq{#2}+\sum_{#3}^{#4}\susyq{i}} }
	{(\mhvdenom{#2}{#1}{#3}{#4})^2} }
\nc\susyq[1]{\ang{#1}\eta_{#1}}
\begin{document}

\title{Supersymmetric Yang-Mills and Supergravity Amplitudes at One Loop}
\author{Anthony Hall}
\affiliation{{} Department of Physics and Astronomy, UCLA\\
\hbox{Los Angeles, CA 90095--1547, USA}
\\{\tt anthall@physics.ucla.edu}
}

\date{\today}

\begin{abstract}
By applying the known expressions for SYM and SUGRA tree amplitudes, we write generating functions for the NNMHV box coefficients of SYM as well as the MHV, NMHV, and NNMHV box coefficients for SUGRA.  The all-multiplicity generating functions utilize covariant, on-shell superspace whereby the contribution from arbitrary external states in the supermultiplet can be extracted by Grassmann operators.  In support of the relation between dual Wilson loops and SYM scattering amplitudes at weak coupling, the SYM amplitudes are presented in a manifestly dual superconformal form.  We introduce ordered box coefficients for calculating SUGRA quadruple cuts and prove that ordered coefficients generate physical cut amplitudes after summing over permutations of the external legs.  The ordered box coefficients are produced by sewing ordered  subamplitudes, previously used in applying on-shell recursion relations at tree level.  We describe our verification of the results against the literature, and a formula for extracting the  contributions from external gluons or gravitons to NNMHV superamplitudes is presented.
\end{abstract}

\maketitle
\tableofcontents
\section{Introduction}

Supersymmetric gauge theory is profoundly linked to string theory, perturbatively and at strong coupling.  The prescient work of Nair \cite{nair} recognized the Parke-Taylor scattering amplitudes of SYM as fermion correlators on a sphere.  This result was generalized by Witten \cite{Witten} to describe a weak-weak coupling duality between SYM and D-instantons of the B-model topological string in supersymmetric twistor space.  Other topological, dual descriptions of SYM have been proposed by Berkovits \cite{berkovits1,berkovits2}, Neitzke and Vafa \cite{neitzkevafa}, and Siegel \cite{siegel}.  The representation of scattering amplitudes in twistor space has been studied since the inception of twistor theory \cite{twistors}.  The BCFW on-shell recursion relations for spacetime signature $(2,2)$ have recently been formulated in twistor space \cite{newtwistors}, and a CSW prescription for SYM is presented in Ref.~\cite{supercsw}.

The AdS/CFT correspondence relates the quantum theories of weakly coupled Type IIB strings in an $AdS_5\times S^5$ geometry to strongly coupled SYM on the four-dimensional boundary of $AdS_5$.  In light of this correspondence, Alday and Maldacena \cite{aldaymaldacena} conjectured that the strong-coupling limit of $n$-gluon scattering amplitudes, to all-loop order, in SYM are related to minimal surfaces in $AdS_5$.  The minimal surface is a polygon with light-like edges $[x_i,x_{i+1}]$, where the dual coordinates $x_i$ are related to the gluon momenta by $p_i^\mu = x_i^\mu-x_{i+1}^\mu$.  This method of evaluating SYM amplitudes is equivalent to calculating a dual Wilson loop along the light-like polygon edges $[x_i,x_{i+1}]$.  The scattering amplitude/dual-Wilson loop duality is conjectured to hold for weak and strong coupling, with agreement confirmed up to the six-point, two-loop MHV amplitude \cite{drummondfourgluon,brandhuberwilson,
drummondloops1,drummondloops2,drummondloops3}.

In Ref.~\cite{drummondmagic}, it was observed that the integrals required for the three-loop MHV amplitudes, calculated by Bern, Dixon, and Smirnov \cite{bernloopiteration}, are conformally covariant when formulated in the dual $x$-coordinates.  The relationship with Wilson loops, which have a conformal symmetry, hinted at the presence of an unexpected dual-conformal symmetry for SYM scattering amplitudes.  The conformal symmetry of Wilson loops is manifested as an ultraviolet-anomalous Ward identity.  The conformal Ward identity dictates the form of the finite part of up-to-five cusp Wilson loops at weak \cite{drummondloops1, drummondloops2} and strong coupling \cite{aldaymaldacenanobds, komargodski}.  Aiming to explain why the MHV amplitudes continue to agree with the Wilson loop duality beyond five cusps, the authors of Ref.~\cite{drummondsym} postulate a new, larger symmetry at work, the $\cal{N}=4$ superconformal symmetry $SU(2,2|4)$ acting on dual superspace coordinates.  

The unitarity method \cite{bernunitarity,bernfusing} supplies the technology we use for manifestly on-shell calculations of loop-diagram quantum corrections to scattering amplitudes in quantum field theory.  In a generalized unitary approach \cite{berngenunitarity}, the coefficients of loop integrals are obtained by ``cutting" multiple virtual particles, exposing the loop-integral coefficients as products of on-shell tree amplitudes.  Quadruple cuts \cite{BCFUnitarity} yield integral coefficients which are the product of four on-shell trees, and quadruple cuts freeze the remaining Lorentz-invariant phase-space integrals to a finite set of solutions for the on-shell loop momenta.   The ``no-triangle property" of SYM and SUGRA \cite{arkanihamed,berngraviton,berntwistorgravity,gravitynmhv,notriangle, trianglecancel} allows one-loop amplitudes to be completely specified by quadruple cuts.  The first cut calculation of SUGRA amplitudes by exploiting the KLT relations between gravity and squared gauge-theory tree amplitudes were carried out in Ref.~\cite{berngaugegravity}.  Multi-leg results for SUGRA box coefficients were first presented in Ref.~\cite{berngraviton}.

On-shell recursion relations for amplitudes in on-shell superspace \cite{freedman,brandhubersuperconformal,arkanihamed,freedmanrecursion}  allowed the authors of Ref.~\cite{drummondtrees} to expose SYM tree amplitudes in a manifestly dual superconformal form.  Although infrared divergences spoil the dual conformal properties at one loop, in Ref.~\cite{drummondloop} the authors develop a supersymmetric version of generalized unitarity.  The ratio between the NMHV and MHV one-loop superamplitude is a dual conformal invariant \cite{drummondloop,brandhuberdualconf,freedmandualconf}.  The use of covariant, on-shell superspace allows the supersymmetric sums over states crossing unitarity cuts to be written as Grassmann integrals \cite{drummondsym}.  Diagrammatic methods for directly computing such sums are presented in Ref.~\cite{bernsupersums}.  
In the present paper, we apply generalized unitarity in on-shell superspace to calculate the NNMHV amplitudes for SYM at one loop.

The on-shell superspace description of maximally supersymmetric Yang-Mills requires only minor modifications to be applied for SUGRA.  The contributions of MHV and non-MHV amplitudes to SUGRA scattering are likewise classified as coefficients of Grassmann-valued polynomials.  In Ref.~\cite{drummondsugra} the authors invented ``ordered subamplitudes" for SUGRA tree amplitudes.  The subamplitudes are added together with permutations of $(n-2)$ of the external legs to yield a physical amplitude.  The use of ordered subamplitudes allowed efficient application of on-shell recursion relations, and the authors present the MHV, NMHV and NNMHV contributions to SUGRA tree amplitudes.  Echoes of the intriguing squaring relationship between gauge theory and gravity \cite{berngaugegravity,gaugegravity,freedmanmhv} are observed, since the on-shell recursion relations are seeded with MHV and $\mhvbarthree$ amplitudes which are both proportional to squared SYM tree amplitudes.  

In this paper we present the planar, one-loop contributions to $n$-point NNMHV scattering amplitudes in $\cal{N}=4$ Super Yang-Mills (SYM) and $\cal{N}=8$ Supergravity (SUGRA) theories.  Generalized unitarity allows us to utilize the compact representations of tree level scattering amplitudes obtained previously through the use of on-shell recursion relations \cite{drummondtrees,drummondsugra}.  Scattering amplitudes for SYM at weak and strong coupling are conjectured to possess dual superconformal symmetry, a new symmetry beyond the familiar supersymmetry and conformal invariance.  Our results for the one-loop amplitudes of SYM confirm that the NNMHV box coefficients are covariant under dual superconformal transformations.

We prove that ordered tree amplitudes for SUGRA may be sewn together to yield ``ordered box coefficients" via generalized unitarity.  The ordered box coefficients yield physical box coefficients after adding the permutations of all external legs.  We calculate explicit expressions for the ordered box coefficients which contribute to SUGRA at one loop.

This paper is organized as follows.
We begin with a review of the on-shell, covariant superspace formalism for describing scattering amplitudes in SYM and SUGRA.  In this framework the contributions to SUSY amplitudes of Grassmann degree $(\cal{N} k)$ generate the $\nxmhv{k}$ scattering amplitudes.  Next we describe the tree-level SYM amplitudes we require and review generalized unitarity in the construction of supersymmetric box coefficients.  We calculate the NNMHV box diagrams for SYM and present the box coefficients.  Our results for supersymmetric scattering amplitudes are expressed as Grassmann-valued generating functions, exploiting the on-shell superspace formulated by Drummond, Henn, Korchemsky, and Sokatchev \cite{drummondsym}.

In order to efficiently calculate SUGRA box coefficients, we introduce ``ordered box coefficients."  The ordered box coefficients are formed by fusing ordered tree-level subamplitudes via unitarity.  After summing over external leg permutations, the ordered box coefficients yield physical quadruple-cut coefficients.  We proceed then to write the MHV, NMHV, NNMHV box coefficients for SUGRA in the on-shell superspace language.  We then present a simple formula for extracting gluon and graviton scattering amplitudes from the NNMHV superamplitudes.  Finally we describe the checks we have performed in comparison with amplitudes in the literature.


\section{Preliminaries}
\subsection{Spinor helicity formalism}
In order to efficiently utilize the four-dimensional polarization and momenta data for a scattering process, we describe amplitudes in the spinor helicity formalism for massless particles.   In this formalism we write the Weyl spinors $\lambda_\alpha$ and $\tilde\lambda_{\dot\alpha}$ for a particle with complex and null momentum $p$ as
\begin{align}
	\lambda(p) = \ang{p^-} = \ang{p}, \qquad
	\tilde\lambda(p) &= \ang{p^+} = \bra{p}, 
\end{align} 
suppressing the spinor indices.
The convention we use is that $\ang{p}$ and $\bra{p}$ have helicity weights of $\null\mp \frac{1}{2}$, respectively.  This helicity assignment is consistent with the polarization vectors for an on-shell particle with momentum $p$, with reference spinors $\ang{\mu}$ and $\bra{\mu}$,
\begin{align}
	\epslash^+(p) = \frac{\ang{\mu} \bbra{p}}{\spa{\mu}{p}}, 
	\qquad
	\epslash^-(p) = \frac{\ang{p} \bbra{\mu}}{\spb{p}{\mu}}.
\end{align}
The null momentum $\s{p}$ is written in this formalism as a bi-spinor,
\begin{align}
	\s{p} = \lambda(p) \tilde\lambda(p) =  \ang{p}\bbra{p}.
\end{align}
The Lorentz invariant spinor inner-products between spinors for the particles labeled $i$ and $j$ are denoted  
\begin{align}
	\epsilon^{\alpha\,\beta} 
	\lambda_{i\,\alpha} \lambda_{j\,\beta} 
	= \spa{i}{j}, \qquad
	\epsilon^{\dot\alpha\,\dot\beta} 
	\tilde\lambda_{i\,\dot\alpha} \tilde\lambda_{j\,\dot\beta} 
	= \spb{i}{j}.
\end{align}
In this formalism the spinor products for strings of momenta and spinors are expressed as, for example,
\begin{align}
	\bang{a}\s{k}\s{p} \ang{b} = \bang{a} k p \ang{b} = 
	\spa{a}{k}\spb{k}{p}\spa{p}{b}.
\end{align}

\subsection{Covariant description of on-shell superspace}
Now we discuss the manifestly Lorentz covariant description of the $\cal{N}=4$ multiplet of massless states as formulated by  Drummond, Henn, Korchemsky, and Sokatchev \cite{drummondsym}.  The covariant description utilizes the bi-spinor representation of a complex and null momentum in four dimensions, 
\begin{align}
	(\sigma_\mu)_{\alpha \dot\alpha} p^\mu 
	= \pslash_{\alpha \dot\alpha} =
	\lambda_\alpha \tilde\lambda_{\dot\alpha}.
\end{align}
Then the supersymmetry algebra generated by $q^A_\alpha$ and $\bar q_{A \dot\alpha}$ for $1\leq A \leq \cal{N}$ is written as  
\begin{align} \label{susy}
	\{q^A_\alpha, \bar q_{B \dot\alpha}\} 
	= \delta^A_B \, \lambda_\alpha \tilde\lambda_{\dot\alpha}.
\end{align}
In reference to the null momentum $p$, the spinor components of $q^A_\alpha$ can be decomposed into two linearly independent spinors, 
$q^A_\alpha = (q^A_\alpha)_\parallel + (q^A_\alpha)_\perp$, one parallel and one orthogonal to the spinor $\lambda_\alpha$.  The parallel component satisfies $\lambda^\alpha (q^A_\alpha)_\parallel =0$, and one defines the operator $q^A$ through the relation $(q^A_\alpha)_\parallel = \lambda_\alpha q^A$.  Then the operator-valued part of $(q^A_\alpha)_\perp$ is chosen to be $q^A_\perp = \lambda^\alpha q^A_\alpha$.   Similar considerations apply for the decomposition of $\bar q_{A \dot\alpha}$ relative to the spinor $\tilde\lambda_{\dot\alpha}$.  Forming the spinor product with $\lambda$ and $\tilde\lambda$ on both sides of the supersymmetry algebra \refeqn{susy}, one finds the algebra    
\begin{align}\label{covariantalgebra}  \nn
&	\{q^A_\perp, \bar q_{B \perp}\} =
	\{q^A_\perp, \bar q_{B}\} =  
	\{q^A, \bar q_{B \perp}\} = 0, 
\\	&\{q^A, \bar q_{B} \} = \delta^A_B.
\end{align}
Then we can identify a maximal, mutually anticommuting set of operators (i.e.~annihilation operators) to be either 
$\{q^A,\, q^A_\perp,\, \bar q_{A \perp}\}$ 
or 
$\{\bar q_{A},\, q^A_\perp,\, \bar q_{A \perp}\}$.  

The vacuum state is defined as the state annihilated by the chosen set of annihilation operators and with helicity $(h)$ with respect to the null vector $p$.  Following the convention established in Ref.~\cite{drummondsym}, we choose the set of annihilation operators 
$\{q^A,\, q^A_\perp,\, \bar q_{A \perp}\}$.  The remaining operators $\bar q_A$ are creation operators whose action changes the helicity of a state by $(\null-1/2)$.  Then $q^A$ must carry helicity $(\null+1/2)$ to be consistent with the relations \refeqn{covariantalgebra}.  With a vacuum state of helicity $(\null+1)$, the states created by repeated application of $\bar q_A$ have helicities ranging from $(\null-1\leq h \leq 1)$ and produce the $\cal{N}=4$ multiplet of massless states, self-conjugate under CPT.  

The algebra \refeqn{covariantalgebra} is conveniently realized by  Grassmann variables $\eta^A$ satisfying anticommutation relations $\{\eta^A, \eta^B\} = 0$. The operators are identified with 
\begin{align}
	q^A = \eta^A, 
	\quad \bar q_A=\frac{\partial}{\partial \eta^A}.
\end{align}
The Grassmann variables $\eta^A$ transform according to the fundamental representation of the global $SU(4)$ $R$-symmetry group of $\cal{N}=4$ SYM.

All the component states of the on-shell supermultiplet can be assembled into a single super-wavefunction,
\begin{align} \label{superfield} \nn
	\Phi(p,\eta)
 		&= G^+(p) + \eta^A \Gamma_A(p)
		+ \frac{1}{2}  \eta^A \eta^B S_{AB}(p)
		+\frac{1}{3!} \epsilon_{ABCD} 
			\eta^A \eta^B \eta^C \bar{\Gamma}_D
\\ & \qquad
		+\frac{1}{4!} \epsilon_{ABCD}
			\eta^A \eta^B \eta^C \eta^D G^-(p).
\end{align}
The different states in the supermultiplet are obtained from the super-wavefunction by the action of $\bar q_A$. For example, the positive-helicity gluon state is 
$G^+(p)=\Phi(p,\eta)\arrowvert_{\eta=0}$, 
and the negative-helicity gluon state is given by
$G^-(p)=\epsilon^{ABCD} \bar q_A \bar q_B \bar q_C \bar q_D \Phi(p,\eta)\arrowvert_{\eta=0}$.  Each component particle of the supermultiplet is distinguished by a unique power of $\eta$.  Because the Grassmann variables $\eta^A$ have helicity $(\null+1/2)$, each term in the super-wavefunction of \refeqn{superfield} has total helicity $(\null+1)$.  The momentum $p_\mu$ is chosen by convention to be an outgoing momentum for scattering processes.  A physical particle's momentum is null and future pointing.

An exercise in Grassmann integration yields an important identity, the single-particle completeness relation, 
\begin{align} \label{completeness} \nn
	\int d^4\eta \, \Phi(p,\eta) \Phi(-p,\eta) &= 
	G^+(p)G^-(-p)+G^-(p)G^+(-p)+\Gamma_A(p)\bar\Gamma^A(-p)
\\ & \qquad
	+\bar\Gamma^A(p)\Gamma_A(-p) 
	+\frac{1}{2}S^{AB}(p)\bar S_{AB}(-p).
\end{align}
We will apply this identity to write the sum over each on-shell state in the multiplet as a Grassmann integral.  Thus the discrete sum over particle states which cross an on-shell line in unitarity cuts and on-shell recursion relations is replaced by integration.

The SYM scattering amplitudes for $n$ external superparticles are denoted  
\begin{align}
	\cal{A}_n(p_i,\eta_i) = \cal{A}(\Phi_1,\ldots,\Phi_n).
\end{align}
The superamplitude is a generating function for the scattering of all particles in the $\cal{N}=4$ multiplet, since the contribution of various component fields to a superamplitude are distinguished by the particular Grassmann-valued coefficients appearing in the superfield, \refeqn{superfield}.  For example, a gluon MHV amplitude appears as 
\begin{align}
	\cal{A}_n(p_i,\eta_i) = 
			\left( 
			{\eta_1}^1 {\eta_1}^2 
				{\eta_1}^3 {\eta_1}^4 \right)
			\left( 
			{\eta_2}^1 {\eta_2}^2 
				{\eta_2}^3 {\eta_2}^4 \right)
		A_n(1^-,2^-,3^+,\ldots,n^+) + \ldots .
\end{align}
As discussed in \cits{freedmanmhv,freedman}, the component particle scattering amplitudes are obtained by applying Grassmann-variable derivatives to a superamplitude.  Equivalently, as in Ref.~\cite{drummondtrees}, Grassmann integrations can be used to isolate a component scattering amplitude.  Noting that
\begin{align}
	\int d^4\eta\, \eta^1 \eta^2 \eta^3 \eta^4 = 1, 
	\quad \textrm{i.e.} \quad 
	\deltfour{\eta^A} = \frac{1}{4!} \epsilon_{ABCD}
			\eta^A \eta^B \eta^C \eta^D,
\end{align}
and referring to the component fields in \refeqn{superfield}, negative-helicity gluon contributions, for example, to an amplitude are selected by the Grassmann-integration $\int d^4\eta$ while positive-helicity gluons are indicated by a factor of unity.  Thus MHV, NMHV, and NNMHV gluon amplitudes are given, respectively, by 
\begin{align} \label{selectgluons} \nn
	A_n(1^-,2^-,3^+,\ldots,n^+) &= 
		\int d^4\eta_1 d^4\eta_2 \cal{A}_n(p_i,\eta_i), 
\\ \nn
	A_n(1^-,2^-,3^-,4^+,\ldots,n^+) &= 
		\int d^4\eta_1 d^4\eta_2 d\eta_3 \cal{A}_n(p_i,
			\eta_i), 
\\
	A_n(1^-,2^-,3^-,4^-,5^+,\ldots,n^+) &= 
		\int d^4\eta_1 d^4\eta_2 d^4\eta_3 d^4\eta_4 
			\cal{A}_n(p_i,\eta_i).
\end{align}

Because the $\cal{N}=4$ multiplet is CPT self-conjugate, the same on-shell supermultiplet could have been obtained by using the triplet of annihilation operators 
$\{\bar q_{A},\, q^A_\perp,\, \bar q_{A \perp}\}$ with the creation operators $q^A$.  In that case we would write 
\begin{align}
	\bar q_{A} = \bar \eta_A,
\quad
	q^A = \frac{\partial}{\partial \bar \eta_A}.
\end{align}
The Grassmann variables $\bar \eta_A$ transform in the anti-fundamental representation of $SU(4)$.  The conjugate super-wavefunction is
\begin{align} \nn
	\bar\Phi(p,\bar\eta) &= G^-(p) + \bar\eta_A \bar\Gamma^A(p)
		+ \frac{1}{2}  \bar\eta_A \bar\eta_B \bar S^{AB}(p)
		+\frac{1}{3!} \epsilon^{ABCD} 
			\bar\eta_A \bar\eta_B \bar\eta_C \Gamma_D
\\ & \quad
		+\frac{1}{4!} \epsilon^{ABCD}
			\bar\eta_A \bar\eta_B 
			\bar\eta_C \bar\eta_D G^+(p).
\end{align}
For complex momentum, the super-wavefunctions $\Phi(p,\eta)$ and $\bar\Phi(p,\bar\eta)$ are related by a Grassman-variable Fourier transform, 
\begin{align}
	\bar\Phi(p,\bar\eta) = \int d^4\eta e^{\eta^A \bar\eta_A} 
		\Phi(p,\eta).
\end{align}
The conjugate description $\bar{\cal{A}_n}$ of a superamplitude $\cal{A}_n$ may likewise be used, where the conjugate is obtained by the replacements $\lambda\rightarrow\tilde\lambda, \tilde\lambda\rightarrow\lambda,$ and $\eta\rightarrow\bar\eta$,
\begin{align}
	\bar{\cal{A}_n}(\lambda, \tilde\lambda, \eta)
	= \cal{A}_n(\tilde\lambda,\lambda,\bar\eta).
\end{align}
A pair of conjugate superamplitudes are related by a Grassmann Fourier transform,
\begin{align}
	\cal{A}_n(\lambda,\bar\lambda,\eta) 
	= \prod_{i=1}^n \int d^4\bar\eta_i e^{-\eta_i\cdot\bar\eta_i}
	\bar{\cal{A}_n}(\lambda, \tilde\lambda, \eta).
\end{align}
All scattering amplitudes in this paper are given in the ``holomorphic" description, where every particle is described by the super-wavefunction $\Phi(p,\eta)$.

An $n$-point $\cal{N}=4$ Super Yang-Mills amplitude $\cal{A}_n(\lambda,\tilde{\lambda},\eta)$ is invariant under the supersymmetry algebra  generated by 
\begin{align}
	q_\alpha^A = \sum_{i=1}^n \lambda_{i\alpha} \eta_i^A,
	\qquad
	\bar{q}_{\dot{\alpha}A} = \sum_{i=1}^n 
		\tilde{\lambda}_{i\dot	{\alpha}} 
		\frac{\partial}{\partial \eta_i^A},
	\qquad
	\{q_\alpha^A, \bar{q}_{\dot{\alpha}B} \}
		=\delta_B^A \, p_{\alpha \dot{\alpha}},
\end{align}
where $1\leq A \leq 4$ and the total momentum is $p_{\alpha \dot{\alpha}}=\sum_{i=1}^n (p_{i}) _{\alpha \dot{\alpha}}$.  The superamplitude $\cal{A}_n$ can be expressed, for $n\geq4$, as
\begin{align} \label{superamplitude}
	\cal{A}_n(\lambda,\tilde{\lambda},\eta) =  
		\deltfour{p_{\alpha\dot{\alpha}}}
		\delteight{q_{\alpha}^A} 
		\cal{P}_n(\lambda,\tilde{\lambda},\eta),
\end{align}
where $\bar{q}$-supersymmetry requires that $\bar{q} \cal{P}_n = 0$.  For $n\geq4$ superamplitudes,  invariance under $q$-supersymmetry is manifest because of the delta function. The exceptional three-point amplitudes are shown in detail below. 

The $SU(4)$ $R$-symmetry for SYM implies that the superamplitude is an $SU(4)$ singlet.  Then $\cal{P}_n$ is expanded in a series of $SU(4)$-invariant, homogeneous polynomials of degree $(4k)$ in the $\eta$'s,
\begin{align}
	\cal{P}_n = \sum_{k=0}^{n-4} \cal{P}_n^{4k}.
\end{align}
The invariance under the $\bar{q}$ supersymmetry can be used to set to zero two of the $\eta$ variables, corresponding to two external particles in $\cal{A}_n$, so that the total Grassmann degree of the polynomial $\cal{P}_n$ is $(4n-16)$.  Thus the Grassmann degree of the superamplitude, including the factor of $\delteight{q}$,  is $(4n-8)$.

Next we will identify the $\nxmhv{k}$ scattering amplitudes of the component particles in the supermultiplet with the Grassmann polynomials $\cal{P}_n^{4k}$.  With each superparticle carrying total helicity $(\null+1)$, the total helicity of an $n$-point superamplitude is $(\null+n)$.  The scattering amplitude in \refeqn{superamplitude} is the product of a momentum delta function with zero Grassmann-variable and spinor helicity, and the supercharge delta function has Grassmann-variable helicity of $(\null+4)$ and spinor helicity of $(\null-4)$.  Since the momentum and supercharge delta functions in \refeqn{superamplitude} carry total helicity zero, each Grassmann polynomial $\cal{P}_n^{4k}$ has Grassmann-variable helicity $(\null+2k)$ and thus spinor helicity $(n-2k)$.  

The scattering amplitudes for the component particles of the supermultiplet are obtained as coefficients of Grassmann-polynomial factors in the superamplitude.  According to the above helicity count, the component-particle amplitudes arising from $\cal{P}_n^{4k}$ in this way have spinor helicity $(n-2k-4)$.  In other words, the Grassmann polynomial $\cal{P}_n^{4k}$ yields a generating function for the $\nxmhv{k}$ amplitudes, 
\begin{align}
	\cal{A}_n(\nxmhv{k}) =  
		\deltfour{p_{\alpha\dot{\alpha}}}
		\delteight{q_{\alpha}^A} 
		\cal{P}_n^{4k}(\lambda,\tilde{\lambda},\eta).
\end{align}
We will use the following less-concise but simpler notation,
\begin{align}
	\cal{P}_n^{4k} = \cal{P}_n(\nxmhv{k}), 
\end{align}
to make the relationship between Grassmann polynomials and component amplitudes explicit.

Momentum conservation for three-point vertices, $\ang{i}\bbra{i}+\ang{j}\bbra{j}+\ang{k}\bbra{k}=0$ places exceptional constraints on the particles' spinors.  By contracting the momentum conservation condition with, say, $\bra{k}$, we have
\begin{align}\label{threeang}
	\ang{i}\spb{i}{k} + \ang{j}\spb{j}{k} = 0,
\end{align}
and similarly for contractions with $\bra{i}$ and $\bra{j}$.
Instead contracting with $\ang{j}$ we find
\begin{align}\label{threebra}
	\bra{i}\spa{i}{j} + \bra{k}\spa{k}{j} = 0,
\end{align}
and similarly for contractions with $\ang{i}$ and $\ang{k}$.
Both sets of conditions taken together amount to the trivial solution where all momenta vanish.  This is because \refeqn{threeang} and its companions imply that $\ang{i}$, $\ang{j}$, and $\ang{k}$ are all proportional, which, considered together with \refeqn{threebra} and its companions, would imply the vanishing of $\bra{i}$, $\bra{j}$, and $\bra{k}$.  Therefore we must choose one set of solutions, \refeqn{threeang} and its companions or \refeqn{threebra} and its companions, at a three-point vertex.

\section{Scattering Amplitudes for $\cal{N}=4$ SYM}
\subsection{SYM tree amplitudes}
\label{subsection:symtrees}
The tree-level MHV amplitudes of SYM are given by the generating function of Nair, presented in Ref.~\cite{nair}.  The corresponding  Grassmann polynomial of degree zero is 
\begin{align}\label{mhvsym}
	\cal{P}_{n;0}(\mhvt) = \prod_{i=1}^n \frac{1}{\spa{i}{i+1}}.
\end{align}
The MHV amplitudes have obvious $q$-supersymmetry because of the $\delteight{q}$ factor, and the $\bar{q}$-supersymmetry follows from momentum conservation.
MHV three-point tree amplitudes are well-defined only for the kinematics of \refeqn{threebra}.

The exceptional three-point $\overline{\textrm{MHV}}$ vertex, the Grassmann-variable Fourier transform of the conjugate three-point MHV vertex, has a Grassmann degree of four, 
\begin{align}\label{mhvbarthreeamp}
	\cal{A}_{3;0}(\overline{\textrm{MHV}}) = 
	\deltfour{p_{\alpha \dot\alpha}}
	\mhvbar{1}{2}{3}.
\end{align}
The $\mhvbarthree$ vertex requires using the kinematic constraints of \refeqn{threeang}.  By virtue of these constraints, 
\begin{align}
	q = \susyq{1}+\susyq{2}+\susyq{3} 
	= \frac{\ang{1}}{\spb{2}{3}}(\deltmhvbar{1}{2}{3}),
\end{align}
which annihilates \refeqn{mhvbarthreeamp} to ensure $q$-supersymmetry.  The $\bar{q}$-supersymmetry follows from applying the Schouten identity,
\begin{align}
	\bra{i}\spb{j}{k}+\bra{k}\spb{i}{j}+\bra{j}\spb{k}{i}=0.
\end{align}

We will be using the formulas for $\nxmhv{k}$ tree amplitudes deduced from the application of on-shell recursion relations.  Here we review the SUSY generalization \cite{brandhubersuperconformal,arkanihamed} of the BCFW recursion relations \cite{bcf,bcfw}.  A superamplitude $\cal{A}_{n;0}$ becomes a meromorphic function of the complex variable $z$ under the shift of external-particle spinors and Grassmann variables,
\begin{align}
	\ang{1(z)} &= \ang{1} - z \ang{n}, \\ \nn
	\bra{n(z)} &= \bra{n} + z \bra{1}, \\ \nn
	\eta_n(z)  &= \eta_n  + z \eta_1.
\end{align}
This complex shift is chosen to preserve overall momentum and the supercharge $q$,
\begin{align}
p_1(z) + p_n(z) = p_1 + p_n, \qquad
\eta_1 \ang{1(z)} + \eta_n(z) \ang{n} = \susyq{1}+\susyq{n}.
\end{align} 
The Feynman diagram representation of scattering amplitudes implies that $\cal{A}_{n;0}(z)$ has simple poles at the values of $z$ which yield an internal line with on-shell momentum $P(z)$, 
\begin{align}
	P(z)^2 = P^2 + z \ab{n}{P}{1} = 0.
\end{align}
The values of the shift parameter $z$ which yield the multi-particle poles are denoted $z_P$.  

Both SYM and SUGRA amplitudes have the remarkable ultraviolet behavior that $\cal{A}_{n;0}(z)$ vanishes as $z\rightarrow \infty$, which implies, for the contour at infinity,  
\begin{align}\oint \frac{\cal{A}_{n;0}(z)}{z} dz = 0.
\end{align}
Then we apply Cauchy's Theorem, deforming the contour to the origin to yield residues for the the multiparticle poles at $z = z_P$ and at $z=0$. Scattering amplitudes factorize at multiparticle poles, and the residue at $z=0$ is simply the desired, unshifted scattering amplitude.  Then we arrive at the BCFW recursion relations in their supersymmetric form, 
\begin{align}
	\cal{A}_{n;0}(z=0) = \sum_P \int d^8\eta 
		\cal{A}_{n;0}^{\textrm{L}}(z_P) \frac{1}{P^2} 
		\cal{A}_{n;0}^{\textrm{R}}(z_P).
\end{align}
The sum over intermediate particle states has been written as an  integration over the internal particle's Grassmann variable.

In \cits{drummondtrees,drummondsugra}, Drummond et al. use the supersymmetric BCFW recursion relations to develop a graphical  algorithm for writing the SYM $\nxmhv{k}$ tree amplitudes.  
All SYM amplitudes are dual superconformal invariant, depending on the dual conformal invariant functions
\begin{align} \label{generalR}
	R_{n;a_1,b_1;\ldots;a_r,b_r;ab}^{l_1,\ldots,l_r} =
	\frac{	
	\spa{a}{a-1} \spa{b}{b-1}
	\deltfour{\bang{\xi}x_{b_r a}x_{ab}\ang{\theta_{b b_r}}
		 +\bang{\xi}x_{b_r b} x_{ba}\ang{\theta_{a b_r}} }
	}
	{x_{ab}^2 
	\bang{\xi}x_{b_r a}x_{ab}\ang{b}
	\bang{\xi}x_{b_r a}x_{ab}\ang{b-1}
	\bang{\xi}x_{b_r b}x_{ba}\ang{a}
	\bang{\xi}x_{b_r b}x_{ba}\ang{a-1}
	},
\end{align}
where 
\begin{align}
\bang{\xi} = \bang{n}x_{n a_1}x_{a_1 b_1}x_{b_1 a_2}x_{a_2b_2}
		\cdots x_{a_r b_r}.
\end{align}
The dual-superspace variables $x$ and $\theta$ are related to superspace momenta and spinors by
\begin{align} \label{dualvariables}
	p_i &= x_i-x_{i+1} \quad \textrm{and} \quad
	\susyq{i} = \ang{\theta_i} - \ang{\theta_{i+1}},
\quad \textrm{thus},
\\ \nn
	x_{ab} &= x_a - x_b = \sum_{i=a}^{b-1} p_i 
			\quad \textrm{and} \quad
	\ang{\theta_{ab}} = \ang{\theta_a}-\ang{\theta_b} = 
				\sum_{i=a}^{b-1} \susyq{i}.
\end{align}

In an $n$-point superamplitude the subscript indices of the dual conformal invariants which appear range over the values $\{2,\ldots,n-1\}$.  When the index $a$ attains the lower limit of its range, the spinor $\ang{a-1}$ is to be modified according to the superscript indices $\{l_1,\ldots,l_r\}$,
\begin{align} \label{modifyR}
	\bang{a-1} \rightarrow \bang{n}x_{n l_1}x_{l_1 l_2}\cdots
				x_{l_{r-1} l_r}.
\end{align}
Dual conformal invariants with no superscripts present require no modification.  We note for later that the dual conformal invariants (and their modified versions) depend on $n$ only through the spinor $\ang{n}$ and have phase weight zero in $\ang{n}$.

For our present purposes we need only the NMHV and NNMHV tree amplitudes.  The dual conformal invariants which appear in these amplitudes are given explicitly by
\begin{align} \label{nnmhvR} \nn
&	R_{n;ab} = 
		\frac{
		\spa{a}{a-1} \spa{b}{b-1}
		\deltfour{ \bang{n} x_{na} x_{ab} \ang{\theta_{bn}} 
		 	 + \bang{n} x_{nb} x_{ba} \ang{\theta_{an}} }
		}
		{ x_{ab}^2
		\bang{n}x_{na}x_{ab}\ang{b}
		\bang{n}x_{na}x_{ab}\ang{b-1} 
		\bang{n}x_{nb}x_{ba}\ang{a}
		\bang{n}x_{nb}x_{ba}\ang{a-1} }, 
\\
&	R_{n;ab;cd} =
	\frac{
	\spa{c}{c-1} \spa{d}{d-1} 
	\deltfour{ 
	\bang{\xi} x_{bc} x_{cd} \ang{\theta_{db}}
	+
	\bang{\xi} x_{bd} x_{dc} \ang{\theta_{cb}} }  
	}
	{x_{cd}^2 
	\bang{\xi} x_{bc} x_{cd} \ang{d}
	\bang{\xi} x_{bc} x_{cd} \ang{d-1}
	\bang{\xi} x_{bd} x_{dc} \ang{c}
	\bang{\xi} x_{bd} x_{dc} \ang{c-1} }.
\end{align}
where $\bang{\xi} = \bang{n} x_{na} x_{ab}$.
The NMHV and NNMHV tree amplitudes in SYM are generated by the Grassmann-valued polynomials 
\begin{align} \label{nnmhvsym}
	\cal{P}_{n;0}(\nmhv) &= 
	\frac{1}{\prod_1^n \spa{i}{i+1}}
	\sum_{2\leq a,b \leq n-1} R_{n;ab}, 
\\ \nn
	\cal{P}_n(\nxmhv{2}) &=
	\frac{1}{\prod_1^n \spa{i}{i+1}}
	\sum_{2\leq a,b\leq n-1} R_{n;ab} 
	\left[
	\sum_{a\leq c,d<b}R_{n;ab;cd}^{ba}
	+ 
	\sum_{b\leq c,d<n}R_{n;cd}^{ab}
	\right].
\end{align}
We are using a convention for double summations where it is understood that $j \geq i+2$ in a sum $\sum_{i,j}$.

In order to carry out the Grassmann integrals appearing in unitarity cuts, it will be important to notice that the dual conformal invariant factors in the SYM tree amplitudes are independent of the Grassmann variables $\eta_1$ and $\eta_n$. The dual conformal invariant functions all share the property that they depend on the Grassmann variables through the dual superspace coordinates $\theta_{xb_r} = \sum_{i=x}^{b_r}\susyq{i}$,  where the indices range only over the values $2\leq x < b_r \leq n-1$.  In this form of presenting the amplitudes, therefore, the Grassmann variables for the external particles $\eta_1$ and $\eta_n$ appear only in the overall supersymmetric delta function.  

\subsection{SYM one-loop amplitudes}
%
%
%
\begin{figure}
\centering
\includegraphics{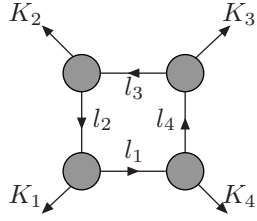}
\caption{The kinematics used for the box coefficients.  All external legs are outgoing and the loop momenta point counterclockwise.  Gray blobs indicate on-shell tree amplitudes.
\label{boxdiagram}}
\end{figure}

The planar, color-ordered, one-loop amplitudes $\cal{A}_{n;1}$ in SYM can be decomposed onto a basis of scalar box integrals \cite{bernunitarity,bernfusing} with Grassmann-valued box coefficients,
\begin{align} \label{symboxdecomposition}
	\cal{A}_{n;1} = \deltfour{p} \sum_{\textrm{partitions}} 
		\left( 
		\cal{C}^{4m} I^{4m} + \cal{C}^{3m} I^{3m} + 
		\cal{C}^{2mh} I^{2mh} + 
		\cal{C}^{2me} I^{2me} + \cal{C}^{1m} I^{1m}
		 \right).
\end{align}
The summation runs over all possible distributions of the color-ordered external particles, and the dimensionally-regularized scalar box integrals are
\begin{align} \label{scalarintegral}
	I(K_1,K_2,K_3,K_4) = -i (4\pi)^{2-\epsilon}
		\int \frac{d^{4-2\epsilon}l}{(2\pi)^{4-2\epsilon}} 
		\frac{1}{l^2 (l+K_1)^2 (l+K_1+K_2)^2 (l-K_4)^2}.
\end{align}
As illustrated in Fig.~\ref{boxdiagram}, the $K_i$ are sums of the momenta leaving each corner of the box.  The four-mass ($4m$) integrals correspond to $(K_i)^2\neq0$ for all four corners;  three-mass ($3m$) integrals have $(K_i)^2=0$ for exactly one corner;  two adjacent $K_i$ have $(K_i)^2=0$ in the two-mass hard ($2mh$) integral, and two opposite corners have vanishing ${K_i}^2$ in the two-mass easy ($2me$) integral; one-mass ($1m$) integrals have ${K_i}^2=0$ for three corners of the box.  We will frequently use a slight abuse of notation, using the symbol $K_i$ to indicate both the total momentum leaving the $i$-th corner and also the set of external-particle labels for that corner.  For example, we write $K_1 = p_1+\ldots+p_{s-1}$ or $K_1 = \{1,\ldots,s-1\}$ depending on the context.

The Grassmann-valued box coefficients $\cal{C}$ are given by the quadruple cuts of the superamplitude $\cal{A}_{n;1}$,
\begin{align}
	\cal{C}(K_1,K_2,K_3,K_4) = 
		\frac{1}{2}\sum_{\cal{S_{\pm}}}\sum_J
		\prod_{i=1}^4 
		\cal{A^\prime}_{n_i+2;0}(l_i,\{K_i\},-l_{i+1}).
\end{align}
Each $\cal{A^\prime}_{n_i+2;0}(l_i,\{K_i\},-l_{i+1})$ is a SYM tree amplitude, with $n_i$ external particles in the cluster $K_i$, stripped of its momentum-conserving delta function.  The unitarity cut coefficient contains a sum over all the component particles of the supermultiplet which cross each cut loop momentum, distinguished by the spin $J$ of each particle.

There are two solutions $\cal{S}_\pm$ for the complex momenta satisfying the on-shell conditions for the cut loop momenta, $l_i^2=0$.  The pair of general solutions for each loop momentum are given in \cite{BCFUnitarity}.  When one corner of the box is massless,  the two solutions  are given in a simple form by the authors of Ref.~\cite{blackhat}.  For numerically checking our results against seven-point gluon and six-point graviton amplitudes the unitarity cuts with at least one massless corner are sufficient.  Considering the routing of momenta we use in Fig.~\ref{boxdiagram},  the solutions are expressed in terms of the spinors $\ang{1}$ and $\bra{1}$ for the massless corner $K_1$, 
\begin{align} \label{loopsolutions}
(l_1^{(\pm)})^\mu &=  
	-\frac
	{\sandpmmp{1}{\s K_2 \s K_3 \s K_4 \gamma^\mu}{1}}
  	{2 \sandpmmp{1}{\s K_2 \s K_4}{1}} \,,
\quad
(l_2^{(\pm)})^\mu = 
	\frac
  	{\sandpmmp{1}{\gamma^\mu \s K_2 \s K_3 \s K_4}{1}}  
  	{2 \sandpmmp{1}{\s K_2 \s K_4}{1}} \,,\nn \\
(l_3^{(\pm)})^\mu &=  
	-\frac
	{\sandpmmp{1}{\s K_2 \gamma^\mu  \s K_3 \s K_4}{1}}
  	{2 \sandpmmp{1}{\s K_2 \s K_4}{1}} \,,
\quad 
(l_4^{(\pm)})^\mu = 
	\frac
 	{\sandpmmp{1}{\s K_2 \s K_3 \gamma^\mu  \s K_4}{1}}
  	{2 \sandpmmp{1}{\s K_2 \s K_4}{1}} \,. 
\end{align}
Here we use the spinor notation 
$\bbra{1}=\bang{1^-},\, \bang{1}=\bang{1^+}$.
Noticing that the two solutions are distinguished according to whether $\ang{l_1}\propto\ang{l_2}$ or $\bra{l_1}\propto\bra{l_2}$, we see that the kinematic solution $\cal{S}^+$ is applicable when 
$\cal{A}(l_1,1,-l_2)$ is an $\mhvbarthree$ vertex and $\cal{S}^-$ is used for a MHV three-vertex.  In the equations for box coefficients that follow, we leave implicit the sum over appropriate loop momenta solutions.

The contribution from each on-shell particle in the supermultiplet  which crosses a unitarity cut is conveniently calculated by an integral over the Grassmann variable of each superparticle, as indicated by the completeness relation of \refeqn{completeness}. Then the unitarity cuts contributing to a box coefficient take the general form
\begin{align} \label{generalbox}
	\cal{C}(K_1,K_2,K_3,K_4) = \prod_{i=1}^4 \int d\eta_{l_i}
	\cal{A^\prime}_{n_i+2;0}(l_i,\{K_i\},-l_{i+1}).
\end{align}
In the case of $n\geq4$ superamplitudes as given by \refeqn{superamplitude}, the loop Grassmann variables will appear in delta functions of the form $\mhvnumsum{l_{i+1}}{l_i}{a}{b}{j}$ for the cluster of external legs $K_i=p_a+\ldots+p_b$ at the corner of the box.  To carry out the Grassmann integrations which appear in a box coefficient, \refeqn{generalbox}, we will apply the identity
\begin{align} \label{deltaeightfactor} \nn
&	\mhvnumsum{l_{i}}{l_{i+1}}{a}{b}{j} 
\\
& \qquad = 
		\delteightfactorsum{l_{i}}{l_{i+1}}{a}{b}{j}.
\end{align}
The pair of Grassmann delta functions simply freezes the value of the loop variables $\eta_{l_i}$ and $\eta_{l_{i+1}}$.

Certain configurations of on-shell three-vertices which could appear in the unitarity cuts are forbidden because of the kinematic constraints of eqns.~\eqref{threeang} and \eqref{threebra}.  If two on-shell MHV, or $\overline{\textrm{MHV}}$, three vertices are adjacent and thus share a common particle line, the special kinematic constraints would require that the pair of external particles at these vertices must have spinors $\tilde\lambda$, or $\lambda$, respectively, which are proportional.  General kinematics does not allow such a restriction.  Quadruple cut diagrams with an $\mhvbarthree$ and $\textrm{MHV}_3$ vertex at opposite corners also vanish for kinematic reasons.
%
%
%
\begin{figure}
\centering
\includegraphics{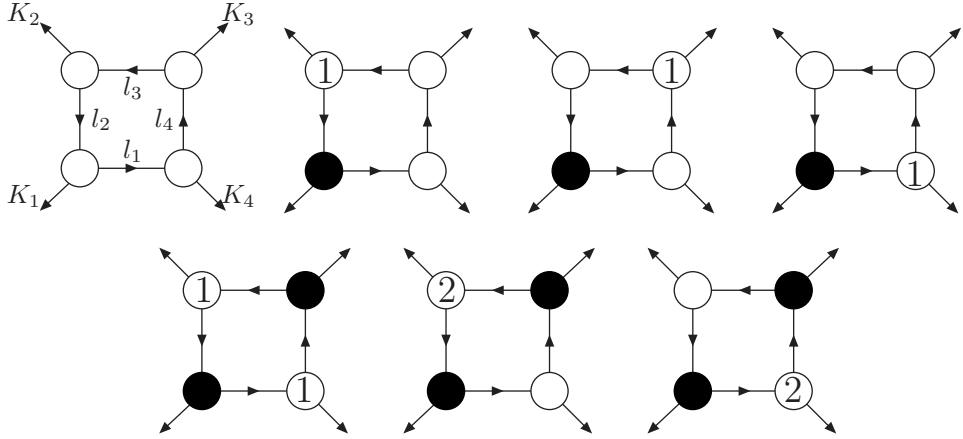}
\caption{The quadruple-cut diagrams which contribute to NNMHV box coefficients.  $\nxmhv{k}$ tree amplitudes are labeled with the number $k$, and $\mhvbarthree$ and MHV tree amplitudes are indicated by black and white blobs, respectively.  The diagrams are indicated in the text with the labels $4m$, $I\!I\!I(A-C)$, and $I\!I\!(A-B)$, respectively.  The last pair of diagrams, which differ by a simple relabeling, are collectively denoted $I\!I(B)$.
\label{boxdiagramnnmhv}}
\end{figure}
%
%

\subsection{NNMHV box coefficients for SYM}
Now we describe all the unitarity cuts which contribute to a NNMHV box coefficient, as in \refeqn{generalbox}.  The NNMHV superamplitudes have total Grassmann degree of $16$, whereas the Grassmann degree of $\mhvbarthree$ vertices, MHV amplitudes, and NMHV  amplitudes are $4$, $8$, and $12$, respectively.  The Grassmann loop integrations for the unitarity cuts each reduce the total Grassmann degree by four, for a net contribution of $-16$.  Thus to have a NNMHV one-loop superamplitude we require the box coefficients to be built from tree amplitudes with total Grassmann degree of $32$.  This is achieved in four different ways;  there can be four MHV tree amplitudes, one NMHV with two MHV and one $\mhvbarthree$ tree amplitude, one NNMHV with one MHV and two $\mhvbarthree$ tree amplitudes, or two NMHV and two $\mhvbarthree$ tree amplitudes.  The kinematic restriction on $\mhvbarthree$ vertices which share a common particle forbids the case with three $\mhvbarthree$ vertices and a $\nxmhv{3}$ tree amplitude.  Fig.~\ref{boxdiagramnnmhv} illustrates all the box diagrams required for the NNMHV one-loop superamplitudes. 

\label{subsection:nnmhvsym}
\subsubsection{All-MHV cut contributions to the NNMHV box coefficients}
The unitarity cuts built from four MHV tree amplitudes have Grassmann degree $16$ and thus contribute to the NNMHV one-loop SYM amplitude.  The result for this cut coefficient is  presented in Ref.~\cite{drummondloop}.  After calculating the four-mass, all-MHV cut contribution we can immediately obtain the all-MHV cut contributions to the three-, two-, and one-mass box coefficients by simply restricting the number of external particles at each tree.  We write $\cal{C}_{r,s,t,u}^{4m}(\nnmhv)$ to denote the only contribution to the four-mass box coefficient, where  
$K_1=p_r+\cdots+p_{s-1}$, $K_2=p_s+\cdots+p_{t-1}$, $K_3=p_t+\cdots+p_{u-1}$, and $K_4=p_u+\cdots+p_{r-1}$, 
and we have 
\begin{align} 
&	\cal{C}_{r,s,t,u}^{4m}(\nnmhv) 
	= \prod_{j=1}^{4} \int d^4\eta_{l_j}  
\\ \nn &
\times
	\mhvn{l_1}{l_2}{r}{s-1}
\times
	\mhvn{l_2}{l_3}{s}{t-1}
\\ \nn &
\times 
	\mhvn{l_3}{l_4}{t}{u-1}
\times
	\mhvn{l_4}{l_1}{u}{r-1}.
\end{align}
The sum of the delta-functions' arguments yields the supersymmetric delta function $\delteight{q}=\delteight{\sum_1^n\susyq{i}}$.  We  replace the argument of the first delta function with $q$ and set the first delta function aside.  Now we use \refeqn{deltaeightfactor} to factor the three remaining $\delta^{(8)}$ functions into pairs of $\delta^{(4)}$ functions to find that their product is
\begin{align} \label{fourmassfactor}
&
	\delteightfactor{l_2}{l_3}{s}{t-1}
\\ \nn 
	\times &
	\delteightfactor{l_3}{l_4}{t}{u-1}
\\ \nn 
	\times &
	\delteightfactor{l_4}{l_1}{u}{r-1}.
\end{align}
The Grassmann integrations over $\eta_{l_2}$ and $\eta_{l_1}$ are now  trivial, and the final integrations simply freeze the values of $\eta_{l_3}$ and $\eta_{l_4}$.  Carrying out the Grassmann integration over the product of delta functions in \refeqn{fourmassfactor} gives
\begin{align} \label{fourmassint} \nn
&
	\frac{1}{\spa{l_3}{l_4}^4} 
	\deltfour{\sum_{t}^{u-1} \eta_i \spa{l_4}{i} \spa{l_3}{l_2}
		 -\sum_{s}^{t-1} \eta_i \spa{l_2}{i} \spa{l_4}{l_3} }
\\ \nn & \qquad \times
	\deltfour{\sum_{u}^{r-1} \eta_i \spa{l_1}{i} \spa{l_4}{l_3}
		 -\sum_{t}^{u-1} \eta_i \spa{l_3}{i} \spa{l_1}{l_4} }
\\ \nn &= 
	\frac{1}{\spa{l_3}{l_4}^4 \spb{l_1}{l_3}^4 \spb{l_2}{l_4}^4 }
	\deltfour{\sum_{t}^{u-1} \eta_i \bang{l_3}l_2 l_4 \ang{i}
		 -\sum_{s}^{t-1} \eta_i \bang{l_3}l_4 l_2 \ang{i} }
\\ \nn & \qquad \times
	\deltfour{\sum_{u}^{r-1} \eta_i \bang{l_4}l_3 l_1 \ang{i} 
		 -\sum_{t}^{u-1} \eta_i \bang{l_4}l_1 l_3 \ang{i} }
\\ \nn &=
	\frac{1}{\spa{l_3}{l_4}^4 \spb{l_1}{l_3}^4 \spb{l_2}{l_4}^4 }
	\deltfour{ \bang{l_3} x_{ts} x_{su} \ang{\theta_{ut}} 
		 + \bang{l_3} x_{tu} x_{us} \ang{\theta_{st}} }
\\  & \qquad \times
	\deltfour{ \bang{l_4} x_{ut} x_{tr} \ang{\theta_{ru}}  
		 + \bang{l_4} x_{ur} x_{rt} \ang{\theta_{tu}} }.
\end{align}
In the second equality of \refeqn{fourmassint} we used the identities 
\begin{align} \label{fourmassterm}
\nn
	\bang{l_3}l_2 l_4 \ang{i} 
	=& 
	\bang{l_3}(l_2-l_3)(l_4-l_2)\ang{i} 
\\ \nn	=& 
 	\bang{l_3}(K_3+K_4+K_1)(K_2+K_3)\ang{i}
\\	=&
	\bang{l_3} x_{ts} x_{su} \ang{i},
\end{align}
and similarly for the other terms in the delta functions, introducing in the last line the dual superspace coordinates from \refeqn{dualvariables}.  The last line of \refeqn{fourmassint} also applies the overall supercharge conservation to write, for example, 
\begin{align} 
	\sum_{t}^{u-1} \eta_i \ang{i} 
	=
	-\sum_{u}^{t-1} \eta_i\ang{i} 
	=
	-\theta_{ut}.
\end{align}
Separating the MHV tree-amplitude factor and simplifying the remaining denominator using \refeqn{fourmassterm}, we find 
\begin{align}
&	\cal{C}_{r,s,t,u}^{4m}(\nnmhv) 
= \nn
	 \frac{\spa{l_4}{l_1} \spa{l_3}{l_2}}
		{\spa{l_2}{l_1} \spa{l_3}{l_4}}
	\times \frac{\delteight{q} }{\prod_1^n \spa{i}{i+1} }
\\ \nn &
\times
	\frac{
		\spa{s}{s-1} \spa{u}{u-1}
		\deltfour{ \bang{l_3} x_{ts} x_{su} \ang{\theta_{ut}} 
		 + \bang{l_3} x_{tu} x_{us} \ang{\theta_{st}} }
		}
		{ \bang{l_3}x_{tu}x_{us}\ang{s}
		\bang{l_3}x_{tu}x_{us}\ang{s-1} 
		\bang{l_3}x_{ts}x_{su}\ang{u}
		\bang{l_3}x_{ts}x_{su}\ang{u-1} }
\\ & 
\times
	\frac{
		\spa{r}{r-1} \spa{t}{t-1}
		\deltfour{ \bang{l_4} x_{ut} x_{tr} \ang{\theta_{ru}} 
		 + \bang{l_4} x_{ur} x_{rt} \ang{\theta_{tu}} }
		}
		{ \bang{l_4}x_{ut}x_{tr}\ang{r}
		\bang{l_4}x_{ut}x_{tr}\ang{r-1} 
		\bang{l_4}x_{ur}x_{rt}\ang{t}
		\bang{l_4}x_{ur}x_{rt}\ang{t-1} }.
\end{align}

In order to simplify the prefactor for four-mass box configurations of momenta, I multiply the numerator and denominator by the factor
$\spb{l_1}{l_2} \spb{l_4}{l_3}$  to find that the numerator is
\begin{align} \nn
	\bbra{l_1}l_2 l_3 l_4 \ang{l_1} &=
	\textrm{tr}_{(-)}
	(\slashvar{l_1}\slashvar{l_2}\slashvar{l_3}\slashvar{l_4}) 
\\ \nn
	&= \frac{1}{2}
	(
	2\lp{l_1}{l_2} \, 2\lp{l_3}{l_4} + 
	2\lp{l_2}{l_3} \, 2\lp{l_1}{l_4} -
	2\lp{l_1}{l_3} \, 2\lp{l_2}{l_4} +
	4 i \epsilon_{\mu\nu\rho\sigma} 
	l_1^\mu l_2^\nu l_3^\rho l_4^\sigma ) 
\\ 
	&= \frac{1}{2}
	( \Delta_{r,s,t,u} +
	4 i \epsilon_{\mu\nu\rho\sigma} 
	l_1^\mu l_2^\nu l_3^\rho l_4^\sigma ).
\end{align}
The factor $\Delta_{r,s,t,u}$ has the expression
\begin{align}
	\Delta_{r,s,t,u} = 
	x_{rs}^2 x_{tu}^2+x_{ru}^2 x_{st}^2-x_{rt}^2 x_{su}^2.
\end{align}
The epsilon tensor piece can be written as 
$\epsilon_{\mu\nu\rho\sigma} l^\mu K_2^\nu K_3^\rho K_4^\sigma$ by applying momentum conservation.  Inside the loop integral, this term vanishes since the loop momentum $l^\mu$ must integrate to a sum of the external momenta $K_i^\mu$.

In cases where the four-mass box degenerates to a three-mass or easy two-mass box, we can encounter $\bra{l_1}\propto\bra{l_2}$ at a MHV three-point vertex.  The factor $\spb{l_1}{l_2}$ vanishes in these cases, and we instead multiply by the factor $\spb{l_1}{l_3} \spb{l_2}{l_4}$.  The spinor products in the numerator can then be written in terms of external leg variables by applying the loop momenta solution $\cal{S}^-$ in \refeqn{loopsolutions}.  The denominator in these cases is 
\begin{align} \nn \label{traceminus}
	\bang{l_2}l_1 l_3 l_4 \bra{l_2} &= 
		\bang{l_2}l_1 l_3 l_4 \bra{l_2} + 
		\bbra{l_2}l_1 l_3 l_4 \ang{l_2} 
\\ \nn
	&= \textrm{tr}
	(\slashvar{l_2}\slashvar{l_1}\slashvar{l_3}\slashvar{l_4}) 
\\
	&=  
	\Delta_{r,r+1,t,u}.
\end{align}
The factor $\Delta$ we use is twice that found in Drummond et al.

We have now obtained the complete four-mass box coefficient, in agreement with the result given in Ref.~\cite{drummondtrees}.  The four-mass box coefficient is 
\begin{align} \label{fourmass}
	\cal{C}_{r,s,t,u}^{4m}(\nnmhv) =
	\frac{ 
	\Delta_{r,s,t,u}}
	{2 x_{rs}^2 \, x_{tu}^2}
	x_{su}^2 x_{rt}^2
	R_{l_3;tsu} R_{l_4;urt} 
	\times \frac{\delteight{q} }{\prod_1^n \spa{i}{i+1} },
\end{align}
where the dual conformal invariant \cite{drummondloop} is
\begin{align}
	R_{l;cab} = 
		\frac{
		\spa{a}{a-1} \spa{b}{b-1}
		\deltfour{ \bang{l} x_{ca} x_{ab} \ang{\theta_{bc}} 
		 	 + \bang{l} x_{cb} x_{ba} \ang{\theta_{ac}} }
		}
		{ x_{ab}^2
		\bang{l}x_{cb}x_{ba}\ang{a}
		\bang{l}x_{cb}x_{ba}\ang{a-1} 
		\bang{l}x_{ca}x_{ab}\ang{b}
		\bang{l}x_{ca}x_{ab}\ang{b-1} },
\end{align}
  The dual conformal invariant $R_{l;cab}$ is identical to $R_{c;ab}$ in \refeqn{nnmhvR}, but with the spinor $\ang{c}$ replaced by the loop momentum spinor $\ang{l}$.
\subsubsection{NNMHV box coefficients with one $\mhvbarthree$ vertex}
The unitarity cuts built from sewing a NMHV tree amplitude, two MHV amplitudes, and one $\mhvbarthree$ amplitude have a total Grassmann degree of $16$ and thus contribute to the NNMHV one-loop superamplitude.  For the three-mass cut contribution we label the external momenta as $K_1 = p_r$ for the external leg at the $\mhvbarthree$ vertex, $K_2 = p_{r+1}+\ldots+p_{s-1}$, $K_3 = p_s+\ldots+p_{t-1}$, and $K_4 = p_t+\ldots+p_{r-1}$.  We calculate the single-$\mhvbarthree$ contribution to the three-mass box coefficient, 
denoted $\cal{C}_{r,r+1,s,t}^{I\!I\!I}(\nnmhv)$.  The two- and one-mass cut contributions from such unitarity cuts are determined by restricting the numbers of external legs at each corner of the box.    

There are three  distinct configurations to consider, as shown in Fig.~\ref{boxdiagramnnmhv}, depending on the placement of the NMHV tree amplitude relative to the $\mhvbarthree$ vertex.  In cases where the cluster of external legs $K_2$ is attached to the NMHV tree, the unitarity cut yields 
\begin{align} 
&	\cal{C}_{r,r+1,s,t}^{I\!I\!I(A)}(\nnmhv) = 
	\prod_{j=1}^{4} \int d^4\eta_{l_j}   
	\mhvbar{l_2}{l_1}{r}
\\  \nn
& \quad \times  
	\mhvn{l_2}{l_3}{r+1}{s-1} \times \sum_{a,b} R_{l_3;ab}
\\  \nn
& \quad \times   
	\mhvn{l_3}{l_4}{s}{t-1}
\times
	\mhvn{l_4}{l_1}{t}{r-1},
\end{align}
where the indices $a$ and $b$ in the NMHV factor satisfy 
$a\geq r+1$ and $a+2\leq b \leq s-1$.
As noted in Section \ref{subsection:symtrees}, the factor $R_{l_3;ab}$ does not depend on the Grassmann variables $\eta_{l_2}$ or $\eta_{l_3}$.  Thus the loop Grassmann integrations only affect the delta functions in the integrand, which explicitly display the loop's Grassmann variables.

The product of the delta functions which appears is exactly the same as in the three-mass box coefficient for NMHV superamplitudes calculated by Drummond et al.  We conclude that 
\begin{align} \label{onemhvbara}
	\cal{C}_{r,r+1,s,t}^{I\!I\!I(A)}(\nnmhv)
	&= \cal{C}_{r,r+1,s,t}^{3m}(\textrm{NMHV})
	\times \sum_{r+1\leq a,b< s} R_{l_3;ab}
\\ \nn 
	&= 
	\Delta_{r,r+1,s,t} 
	R_{r;st} 
	\sum_{r+1\leq a,b< s} R_{l_3;ab}
\times
	\mhv{q}{\prod_1^n \spa{i}{i+1}}.
\end{align}
For this kinematic arrangement, the factor $\Delta_{r,s,t,u}$ simplifies because $x_{r,r+1}^2 = p_r^2 = 0$ to become 
\begin{align}
	\Delta_{r,r+1,s,t} = 
	x_{rt}^2 x_{r+1s}^2 - x_{rs}^2 x_{r+1t}^2.
\end{align}
We have used an identity similar to \refeqn{traceminus}, leading to a result for $\cal{C}_{r,r+1,s,t}^{3m}(\textrm{NMHV})$ which is twice that found in Ref.~\cite{drummondloop}.

The other single-$\mhvbarthree$ contributions to the three-mass box coefficient are illustrated in Fig.~\ref{boxdiagramnnmhv}.  These two diagrams have an NMHV tree amplitude at the corners of the box carrying momenta $K_3$ and $K_4$, respectively, 
\begin{align} \label{onemhvbarbc}
	\cal{C}_{r,r+1,s,t}^{I\!I\!I(B)}(\nnmhv)
	&= \cal{C}_{r,r+1,s,t}^{3m}(\textrm{NMHV})
	\times \sum_{s\leq a,b< t} R_{l_4;ab}
\\ 
	\cal{C}_{r,r+1,s,t}^{I\!I\!I(C)}(\nnmhv)
	&= \cal{C}_{r,r+1,s,t}^{3m}(\textrm{NMHV})
	\times \sum_{t\leq a,b< r} R_{l_1;ab}.
\end{align}
The kinematic constraint 
$\ang{l_1}=-\frac{\spb{r}{l_2}}{\spb{l_1}{l_2}} \ang{r}$ 
from the $\mhvbarthree$ vertex can be used to write $R_{l_1;ab}=R_{r;ab}$.  The factor $R_{r;ab}$ has zero phase weight in $\ang{r}$, so no additional factors are introduced by this replacement.  

Care must be taken when interpreting the dual conformal invariants associated with these box diagrams.  Since the loop leg $l_3$ is adjacent to the external leg labeled $s$, it is easy to show that  
$\bang{l_3}x_{l_3a} = \bang{l_3}x_{sa}$.  Then we can write 
$R_{l_3;ab} = R_{l_3;sab}$ and similarly for the other NMHV factors.  Note also that when the label $(s-1)$ appears in the factor $R_{l_4;ab}$, it refers to the loop leg $l_3$.  Similar considerations apply when expressing the factor $R_{l_1;ab}$.

Finally we have the total contribution to the box coefficients from  diagrams with a single $\mhvbarthree$ vertex, 
\begin{align} \label{onemhvbar} 
	\cal{C}_{r,r+1,s,t}^{I\!I\!I}(\nnmhv) &=
	\cal{C}_{r,r+1,s,t}^{I\!I\!I(A)}+
	\cal{C}_{r,r+1,s,t}^{I\!I\!I(B)}+
	\cal{C}_{r,r+1,s,t}^{I\!I\!I(C)}
\\ \nn
	& = 
	\Delta_{r,r+1,s,t} 
	\mhv{q}{\prod_1^n \spa{i}{i+1}}
\\ \nn 
& \quad \times
	R_{r;st} 
	\left(\sum_{r+1\leq a,b< s} R_{l_3;ab}
		+\sum_{s\leq a,b< t} R_{l_4;ab}
		+\sum_{t\leq a,b< r} R_{r;ab} \right)
\end{align}

\subsubsection{NNMHV box coefficients with two $\mhvbarthree$ vertices}
The last ingredient to complete the NNMHV box coefficients are cut contributions with two $\mhvbarthree$ vertices.  We will calculate the two-mass configurations $\cal{C}_{r,r+1,s,t}^{I\!I}(\nnmhv)$, from which the one-mass contributions will be obtained by restricting the number of external legs at corners of the box.  The kinematic constraint on adjacent $\mhvbarthree$ vertices means that two-mass cut contributions with two $\mhvbarthree$ vertices only contribute to the two-mass easy box coefficients.  Then the cut contribution to NNMHV one-loop superamplitudes with two NMHV tree amplitudes is
\begin{align} \nn
	\cal{C}_{r,r+1,s,s+1}^{I\!I(A)}(\nnmhv) = 
	\prod_{j=1}^{4} \int d^4\eta_{l_j} & 
	\mhvbar{l_2}{l_1}{r}
\\ \nn
\times &
	\mhvn{l_2}{l_3}{r+1}{s-1} \times \sum_{a,b} R_{l_3;ab}
\\ \nn
\times &
	\mhvbar{l_4}{l_3}{s}
\\ 
\times &
	\mhvn{l_4}{l_1}{t}{r-1} \times \sum_{c,d} R_{l_1;cd},
\end{align}
where the indices $a$ and $b$ in the first NMHV factor satisfy 
$a\geq r+1$ and $a+2\leq b \leq s-1$, and in the second factor we have 
$c\geq s+1$ and $c+2\leq d \leq r-1$.  

As in the three-mass box coefficients, the NMHV factors in the integrand $R_{l_3;ab}$ and $R_{l_1;cd}$ are independent of the loop Grassmann variables and are therefore untouched by the Grassmann integrations.  With these dual conformal invariant factors aside, the remaining product of delta functions and denominators is identical to those which appear in the MHV two-mass easy coefficient calculated in Ref.~\cite{drummondtrees}.  Thus we have 
\begin{align} \nn
	\cal{C}_{r,r+1,s,s+1}^{I\!I(A)}(\nnmhv) &=
	\cal{C}_{r,r+1,s,s+1}^{2me}(\mhvt)
	\times 
	\sum_{r+1\leq a,b<s} R_{l_3;ab} 
	\times 
	\sum_{s+1\leq c,d<r} R_{l_1;cd}
\\  &=
	\Delta_{r,r+1,s,s+1} 
	\sum_{r+1\leq a,b<s} R_{s;ab} 
	\sum_{s+1\leq c,d<r} R_{r,cd} 
	\times
	\mhv{q}{\prod_1^n \spa{i}{i+1}} ,
\end{align}
where we have applied the kinematic constraints from the $\mhvbarthree$ vertices to make the replacements $\ang{l_1}\rightarrow\ang{r}$ and $\ang{l_3}\rightarrow\ang{s}$ in the dual superconformal invariants.  We have applied the trace identity \refeqn{traceminus} to find a result for $\cal{C}_{r,r+1,s,s+1}^{2me}(\mhvt)$ which is twice that found in Ref.~\cite{drummondloop}.

The last two diagrams in Fig.~\ref{boxdiagramnnmhv} illustrate the contributions to the two-mass easy box coefficient with a single $\nnmhv$ tree amplitude.  This pair of single-$\nnmhv$ diagrams differ by a reflection of the external particle labels, $r\leftrightarrow s$, and we have  
\begin{align} 
&	\cal{C}_{r,r+1,s,s+1}^{I\!I(B)}(\nnmhv) = 
	\prod_{j=1}^{4} \int d^4\eta_{l_j} 
\\ \nn & \quad  \times
	\mhvbar{l_2}{l_1}{r}
\times 
	\mhvn{l_2}{l_3}{r+1}{s-1} 
\\ \nn
& \quad \times 
	\sum_{r+1\leq a,b< s} R_{l_3;ab} 
	\left[
	\sum_{a\leq c,d< b}R_{l_3;ab;cd}^{ba}
	+
	\sum_{b\leq c,d< s}R_{l_3;cd}^{ab}
	\right]
\\ \nn
& \quad \times 
	\mhvbar{l_4}{l_3}{s}
\times
	\mhvn{l_4}{l_1}{t}{r-1}  
\\ \nn
 &
	\qquad+ (r\leftrightarrow s).
\end{align}
Considering that the dual conformal invariants are independent of $\eta_{l_2}$ and $\eta_{l_3}$ and using the $\mhvbarthree$ vertex constraint to replace $\ang{l_3}\rightarrow\ang{s}$, the Grassmann delta functions and denominators reproduce the MHV two-mass easy coefficient,
\begin{align} \nn
	\cal{C}_{r,r+1,s,s+1}^{I\!I(B)}(\nnmhv) &= 
	\cal{C}_{r,r+1,s,s+1}^{2me}(\mhvt)
\\ \nn
& \quad \times
	\sum_{r+1\leq a,b< s} R_{l_3;ab} 
	\left[
	\sum_{a\leq c,d<b} R_{l_3;ab;cd}^{ba}
	+
	\sum_{b\leq c,d< s} R_{l_3;cd}^{ab}
	\right] 
\\ \nn 
& \qquad	+ (r\leftrightarrow s)
\\ \nn  &=
	\Delta_{r,r+1,s,s+1} 
	\times \mhv{q}{\prod_1^n \spa{i}{i+1}}
\\ \nn
& \quad \times
	\sum_{r+1\leq a,b< s} R_{s;ab} 
	\left[
	\sum_{a\leq c,d<b} R_{s;ab;cd}^{ba}
	+
	\sum_{b\leq c,d< s} R_{s;cd}^{ab}
	\right]  
\\ 
& \qquad	+ (r\leftrightarrow s).
\end{align}

The total contribution to the two-mass box coefficient from box diagrams with a pair of $\mhvbarthree$ vertices is 
\begin{align} \label{twomhvbar}
	\cal{C}_{r,r+1,s,s+1}^{I\!I}(\nnmhv) &=
	\cal{C}_{r,r+1,s,s+1}^{I\!I(A)}(\nnmhv)+
	\cal{C}_{r,r+1,s,s+1}^{I\!I(B)}(\nnmhv)
\\ \nn
	& = 
	\mhv{q}{\prod_1^n \spa{i}{i+1}} 
	\Delta_{r,r+1,s,s+1}
	\Bigg( 
	\sum_{r+1\leq a,b<s} R_{s;ab} 
	\sum_{s+1\leq c,d<r} R_{r,cd} 
\\ \nn
& \quad +
	\sum_{r+1\leq a,b< s} R_{s;ab} 
	\left[
	\sum_{a\leq c,d<b}(R_{s;ab;cd}^{ba})
	+
	\sum_{b\leq c,d< s}(R_{s;cd}^{ab})
	\right]
	 + (r\leftrightarrow s)
	\Bigg),
\end{align}
where the instruction to interchange the labels $r$ and $s$ applies only to the NNMHV tree amplitude factor, the second term in parentheses.  Note that because of the dual-coordinate identity $x_{ab} = -x_{ba}$ we have  
$\Delta_{r,r+1,s,s+1} = \Delta_{s,s+1,r,r+1}$, so this object can be factored outside of the $(r\leftrightarrow s)$ interchange.

\subsubsection{Complete box coefficients for one-loop, NNMHV SYM amplitudes}
The above contributions to box coefficients are organized according to the number of three-point $\mhvbartext$ vertices contributing to a quadruple cut.  The non-$\mhvbarthree$ tree amplitudes in the quadruple cuts of the previous subsections have multi-particle clusters of external particles.  By restricting the number of external particles at the non-$\mhvbarthree$ corners of the box, the four-, three-, and two-mass box coefficients calculated above generate the missing three-, two-, and one-mass box coefficients to complete the $n$-point, one-loop, NNMHV superamplitude.

The only contribution to the four-mass box coefficient comes from the all-MHV quadruple cuts, $\cal{C}_{r,s,t,u}^{4m}$, from \refeqn{fourmass}.  In the box function expansion, \refeqn{symboxdecomposition}, the box integral $I_{r,s,t,u}$ multiplies this coefficient.  

The three-mass box coefficient receives contributions from the quadruple cuts $\cal{C}^{I\!I\!I}$, \refeqn{onemhvbar}, and also the four-mass quadruple cut diagrams with a single external particle at exactly one of the corners.  Then, for the three-mass box coefficient which multiplies the box integral $I_{r,r+1,s,t}$ with $K_1^2 = 0$, we have 
\begin{align}
	\cal{C}_{r,r+1,s,t}^{3m} = 
	\cal{C}_{r,r+1,s,t}^{4m} + 
	\cal{C}_{r,r+1,s,t}^{I\!I\!I}.
\end{align}
Here we dispense with the $(\nnmhv)$ labeling of the cut coefficients.

The two-mass hard box coefficients are obtained from the quadruple cuts $\cal{C}^{I\!I\!I}$ by restricting one of the corners adjacent to the $\mhvbarthree$ to have exactly one external leg.  The four-mass quadruple cut with a pair of adjacent MHV three-point vertices does not contribute to the two-mass hard box coefficient because of the kinematic constraint of \refeqn{threebra}.  Furthermore, massless corners of a box coefficient must be an MHV or $\mhvbartext$ tree amplitude because $\nxmhv{k}$ tree amplitudes with three on-shell particles vanish.  Thus, considering Fig.~\ref{boxdiagramnnmhv}, two-mass hard coefficients which multiply the box integral $I_{r,r+1,s,r-1}$ with $K_1^2=K_4^2=0$ are obtained from $\cal{C}^{I\!I\!I(A)}$ by choosing the massless corner $K_4 = \{r-1\}$. The coefficient $\cal{C}^{I\!I\!I(C)}$ contributes by choosing $K_2 = \{r+1\}$ followed by the relabeling $r\rightarrow r-1$.  The final two-mass hard contributions come from $\cal{C}^{I\!I\!I(B)}$ by choosing either $K_4 = \{r-1\}$ or choosing $K_2 = \{r+1\}$ relabeling $r\rightarrow r-1$.  Altogether, for this two-mass hard box coefficient we have 
\begin{align}
	\cal{C}_{r,r+1,s,r-1}^{2mh} =
	\cal{C}_{r,r+1,s,r-1}^{I\!I\!I(A)} + 
	\cal{C}_{r,r+1,s,r-1}^{I\!I\!I(B)} +
	\cal{C}_{r-1,r,r+1,s}^{I\!I\!I(B)} + 
	\cal{C}_{r-1,r,r+1,s}^{I\!I\!I(C)}.
\end{align}

Next we consider the two-mass easy box coefficients which multiply the box integral $I_{r,r+1,s,s+1}$ with $K_1^2=K_3^2=0$.  Such coefficients are obtained from the $\cal{C}^{4m}$ quadruple cuts by restricting two opposite corners to be massless.  The quadruple cuts $\cal{C}^{I\!I}$ of \refeqn{twomhvbar} directly give appropriate two-mass easy box coefficients.  The $\cal{C}^{I\!I\!I}$ quadruple cuts yield two-mass easy box coefficients by taking the massless corner $K_3 = \{s\}$, but the coefficient $\cal{C}^{I\!I\!I(B)}$ vanishes in this case because of the NMHV tree at the corner with momentum $K_3$.  The two-mass easy contributions from the coefficients $\cal{C}^{I\!I\!I(A)}$ and $\cal{C}^{I\!I\!I(C)}$ also vanish because of the kinematic constraints from the three-point vertices at opposite corners.  Then the two-mass easy box coefficient is 
\begin{align}
	\cal{C}_{r,r+1,s,s+1}^{2me} =
	\cal{C}_{r,r+1,s,s+1}^{4m} + 
	\cal{C}_{r,r+1,s,s+1}^{I\!I}.
\end{align}

The only non-vanishing contributions to the one-mass box coefficient are obtained from the quadruple cuts $\cal{C}^{I\!I\!I}(B)$ and $\cal{C}^{I\!I(B)}$ by restricting all corners to be massless except the corner with a NMHV or NNMHV tree amplitude.  All the other one-mass diagrams vanish because of the kinematic restriction on adjacent MHV or $\mhvbartext$ three vertices, or, in the case of $\cal{C}^{I\!I(A)}$, because on-shell, three-point NMHV amplitudes vanish.  Then the one-mass box coefficient which multiplies the box integral $I_{r-2,r-1,r,r+1}$ with the massive corner $K_4$ is
\begin{align}
	\cal{C}_{r-2,r-1,r,r+1}^{1m} = 
	\cal{C}_{r-1,r,r+1,r-2}^{I\!I\!I(B)} + 
	\cal{C}_{r,r+1,r-2,r-1}^{I\!I(B)}.
\end{align}
This completes the specification of the NNMHV box coefficients for SYM.

\section{Scattering amplitudes for $\cal{N}=8$ SUGRA}
As in the SYM theory reviewed earlier, generating functions for the $\cal{N}=8$ SUGRA theory are given a holomorphic description in terms of the anticommuting Grassmann variables $\eta^A$, for $1\leq A\leq 8$, which transform in the fundamental representation of the $R$-symmetry $SU(8)$.  The component states of on-shell SUGRA appear with unique Grassmann-valued coefficients in the super-wavefunction $\Phi(p,\eta)$.  An $n$-point SUGRA amplitude, the generating function for scattering amplitudes of particles in the supermultiplet, is written as
\begin{align}
	\cal{M}_n(p_i,\eta_i) = \cal{M}(\Phi_1,\ldots,\Phi_n).
\end{align}

The Grassmann integral identity, 
\begin{align}
	\int d^8\eta \,
			\eta^1 \eta^2 \eta^3 \eta^4 
			\eta^5 \eta^6 \eta^7 \eta^8 = 1, 
	\quad \textrm{i.e.} \quad 
	\delteight{\eta^A} = 
			\eta^1 \eta^2 \eta^3 \eta^4 
			\eta^5 \eta^6 \eta^7 \eta^8,
\end{align}
allows scattering amplitudes with external gravitons of negative and positive helicity to be selected by applying $\int d^8 \eta$ and $1$, respectively, to superamplitudes.

A general $n$-point $\cal{N}=8$ SUGRA amplitude is supertranslation invariant and can be written for $n\geq4$ as
\begin{align}\label{sugraampl}
	\cal{M}_n = \deltfour{p_{\alpha\dot{\alpha}}}
		\deltsixteen{q_{\alpha}^A} 
		\cal{R}_n(\lambda,\tilde{\lambda},\eta).
\end{align}
The exceptional three-point amplitudes are shown in detail below.
The superamplitudes of the $\cal{N}=8$ theory conserve the supercharge,  
\begin{align}
	q_{\alpha}^A = \sum_{i=1}^n \lambda_{i,\alpha} \eta_i^A.
\end{align}
As in the SYM theory described previously, $\cal{R}_n$ is expanded in a series of $SU(8)$-invariant, homogeneous polynomials of degree $8k$ in the $\eta$'s,
\begin{align}
	\cal{R}_n = \sum_{k=0}^{n-4} \cal{R}_n^{8k}.
\end{align}
The $\bar{q}$ supersymmetry can be utilized to eliminate two of the $\eta$ variables in $\cal{M}_n$ so that the degree of the superamplitude is $8(n-2)$.  The terms in the superamplitude, \refeqn{sugraampl}, of Grassmann degree $8k+16$ are the generating functions for $\nxmhv{k}$ contributions to SUGRA scattering amplitudes.

\subsection{Tree-level ordered subamplitudes}
In Ref.~\cite{drummondsugra}, the SUGRA recursion relations are applied to efficiently calculate explicit, analytic expressions for all-multiplicity MHV, NMHV, and NNMHV tree amplitudes by introducing ordered gravity subamplitudes $M(1,\ldots,n)$ related to the complete, Bose symmetric physical amplitudes by 
\begin{align} \label{ordered}
	\cal{M}_n = \sum_{\sigma(2,\ldots,n-1)} M(1,\ldots,n).
\end{align}
The ordered subamplitudes are defined through on-shell recursion, starting from $M(1,2,3) = \cal{M}_3$, 
\begin{align} \label{orderedbcfw}
	M(1,\ldots,n) = \sum_{i=3}^{n-1} \int d^8\eta 
			M(\widehat{1},2,\ldots,i-1,\widehat{P})
			\frac{1}{P^2}
			M(-\widehat{P},i,\ldots,\widehat{n}),
\end{align}
where, as in the SYM recursion relations, the cyclic order of external legs is preserved in the factorizations of gravity subamplitudes $M$.  The authors of Ref.~\cite{drummondsugra} prove that the results of recursion relations for ordered subamplitudes match the amplitudes obtained through the SUGRA on-shell recursion relations. 

The MHV three-particle amplitude and its Grassman-variable Fourier conjugate, the $\mhvbarthree$ amplitude, are given by
\begin{align}
	M_3(\mhvt) &= \frac{\deltsixteen{\susyq{1}+\susyq{2}+
					\susyq{3}}}
		{(\spa{1}{2}\spa{2}{3}\spa{3}{1})^2} 
\qquad \textrm{i.e.} \qquad 
\cal{R}_3^0 = \frac{1}{(\spa{1}{2}\spa{2}{3}\spa{3}{1})^2},  
\\ \nn 
	M_3(\mhvbartext) &= \frac{
			\delteight{\deltmhvbar{1}{2}{3}}}
		{(\spb{1}{2} \spb{2}{3} \spb{3}{1})^2}.
\end{align}
Applying the on-shell recursion relations to ordered subamplitudes yields
\begin{align}
	M_{n}(\mhvt) = 
	\frac{1}{\prod_1^n \spa{i}{i+1}^2}
	G^{\mhvt}(1,\ldots,n),
\end{align}
where $G^{\mhvt}(1,2,3)=1$ and otherwise 
\begin{align}
	G^{\mhvt}(1,\ldots,n) = x_{13}^2 \prod_{s=2}^{n-3} 
	\frac{\bang{s}x_{s,s+2} x_{s+2,n} \ang{n}}{\spa{s}{n}}.
\end{align}
Comparison with the SYM amplitudes of eqns.~\eqref{mhvsym} and \eqref{mhvbarthreeamp} indicates that the MHV and $\mhvbarthree$ amplitudes for SUGRA are proportional to the ``squared" MHV and $\mhvbarthree$ amplitudes of SYM, as discussed in Ref.~\cite{drummondsugra}.  It is interesting that ``bonus relations" for SUGRA allow MHV tree amplitudes to be written in a form that require permutations over only $(n-3)$ of the external particles \cite{freedmanmhv,bonusrelations}.

The simple ``squaring" relation between MHV and $\mhvbarthree$ amplitudes in SUGRA and SYM allow the results of on-shell recursion for SYM $\nxmhv{p}$ amplitudes to be recycled in the on-shell recursion relations for ordered gravity subamplitudes.  This procedure is explicitly carried out up to NNMHV amplitudes for SUGRA.  The NMHV ordered subamplitude for SUGRA is
\begin{align}
	M_n(\nmhv) = 
	\frac{1}{\prod_1^n \spa{i}{i+1}^2}
	\sum_{2\leq i,j<n} R_{n;ij}^2G_{n;ij}^{\nmhv}.
\end{align}
The factor $G_{n;ij}^{\nmhv}$ and similar factors for $\nxmhv{p}$ amplitudes are expressed in terms of
\begin{align}
	P_{a_1,\ldots,a_r}^{l,u} &=
	\prod_{k=l}^u 
	\frac{
	\bang{k}x_{k,k+2}x_{k+2,a_1}x_{a_1 a_2}
		\cdots x_{a_{r-1} a_r}\ang{a_r} }
	{ \bang{k}x_{a_1 a_2}\cdots x_{a_{r-1} a_r}\ang{a_r} }, 
\\ \nn 
	Z_{b_1,\ldots,b_l; c_1,\ldots,c_r}^{a_1,\ldots,a_u} &=
	\frac{
	\bang{a_1}x_{a_1 a_2}\cdots x_{a_{u-1} a_u}\ang{a_u} }
	{ \bang{b_1}x_{b_1 b_2} \cdots x_{b_{l-1} b_l} x_{c_1 c_2} 
		\cdots x_{c_{r-1} c_r} \ang{c_r}  }.
\end{align}
Note that these functions fail to be conformal invariants due to  breaks in the chains of labels which appear in $\bang{k}x_{a_1 a_2}$ from $P$ and the $x_{b_{l-1} b_l} x_{c_1 c_2}$ in the denominator of $Z$.
Then defining 
\begin{align}
	f_{n;2b} &= x_{1b}^2 \quad \textrm{and} \quad
\\ \nn	f_{n;ab} &= x_{13}^2 (-Z_{n;a-1}^{n,b,a-1}) P_n^{2,a-2} 
		\textrm{ for } a>2,
\\ \nn G_{n;ab}^L &= -Z_{n;b,a,n}^{n,a+1,b,a,n} P_{b,a,n}^{a,b-3} 
		\quad \textrm{and} 
\\ \nn G_{n;ab}^R &= -Z_{n;b,a,n}^{n,b+1,b,a,n} P_n^{b,n-3}, 
\end{align}
the factor in the NMHV SUGRA subamplitude is 
\begin{align}
	G_{n;ab}^{\nmhv} = f_{n;ab} G_{n;ab}^L G_{n;ab}^R.
\end{align}

The $\nxmhv{2}$ SUGRA subamplitude is 
\begin{align} 
&	M_{n}(\nxmhv{2}) = 
	\frac{1}{\prod_1^n \spa{i}{i+1}^2}
\\ \nn
& \quad \times
	\sum_{2\leq a,b< n} R_{n;ab}^2 
	\left[
	\sum_{a\leq c,d<b}(R_{n;ab;cd}^{ba})^2
	H_{n;ab;cd}^{(1)}+
	\sum_{b\leq c,d<n}(R_{n;cd}^{ab})^2
	H_{n;ab;cd}^{(2)}
	\right],
\end{align}
where 
\begin{align}
	H_{n;ab;cd}^{(1)} &=
	f_{n;ab} G_{n;ab}^R 
	\widetilde{f}_{n;ab;cd} G_{n;ab;cd}^L G_{n;ab;cd}^R,
\\ \nn
	H_{n;ab;cd}^{(2)} &=
	f_{n;ab} G_{n;ab}^L 
	\widehat{f}_{n;ab;cd} G_{n;cd}^L G_{n;cd}^R.
\end{align}
The new ingredients here are the $\widetilde{f}$ in $H^{(1)}$,
\begin{align}
	\widetilde{f}_{n;ab;ad} &= -Z_{n;b,a,n}^{n,b,d,a,n},
\\ \nn
	\widetilde{f}_{n;ab;cd} &= 
	\left(-Z_{n;b,a,n}^{n,b,a+1,a,n}\right)
	\left(-Z_{c-1;b,a,n}^{c-1,d,b,a,n}\right) P_{b,a,n}^{a,c-2}
	\quad \textrm{for } c>a,
\end{align}
the $\widehat{f}$ in $H^{(2)}$,
\begin{align}
	\widehat{f}_{n;ab;bd} &= -Z_{n;b,a,n}^{n,d,b,a,n},
\\ \nn
	\widehat{f}_{n;ab;cd} &= 
	\left(-Z_{n;b,a,n}^{n,b+1,b,a,n}\right)
	\left(-Z_{n;c-1}^{n,d,c-1}\right) P_{n}^{b,c-2}
	\quad \textrm{for } c>b,
\end{align}
and the new $G$-factors are 
\begin{align}
	G_{n;ab;cd}^L &= -Z_{n,a,b;d,c,b,a,n}^{n,a,b,c+1,d,c,b,a,n}
			P_{d,c,b,a,n}^{c,d-3},
\\ \nn
	G_{n;ab;cd}^R &= -Z_{n,a,b;d,c,b,a,n}^{n,a,b,d+1,d,c,b,a,n}
			P_{b,a,n}^{d,n-3}.
\end{align}

In preparation for sewing these SUGRA tree amplitudes together for the unitarity cuts of loop amplitudes, we make several remarks about the phase weight of certain spinors which appear in the factors above.  First, $G^{\mhvt}(1,\ldots,n)$ depends on $n$ only through the spinor $\ang{n}$ and has zero phase weight in that spinor.  The factor $P_{a_1,\ldots,a_r}^{l,u}$ depends on $a_r$ only through the spinor $\ang{a_r}$ and has zero phase weight in that spinor.  Similarly,  $Z_{b_1,\ldots,b_l; c_1,\ldots,c_r}^{a_1,\ldots,a_u}$ depends on $a_1$, $b_1$, $a_u$ and $c_r$ only through the spinors $\ang{a_1}$, $\ang{b_1}$, $\ang{a_u}$, and $\ang{c_r}$.  If $a_1=b_1$ then $Z_{a_1,\ldots,b_l; c_1,\ldots,c_r}^{a_1,\ldots,a_u}$ has zero phase weight in $\ang{a_1}$.  Likewise, if $a_u=c_r$ then  $Z_{b_1,\ldots,b_l; c_1,\ldots,a_u}^{a_1,\ldots,a_u}$ has zero phase weight in $\ang{a_u}$.  Altogether we conclude that both SYM and SUGRA amplitudes hold all the phase weight for particle $n$ only in the Parke-Taylor prefactors, $(\prod_1^n\spa{i}{i+1})^{-1}$.

\subsection{Ordered subamplitudes at one-loop}
The one-loop planar SUGRA amplitudes $\cal{M}_{n;1}$, like the previous SYM amplitudes, are known to depend only on box integral functions.  Then the one-loop planar SUGRA amplitudes have a scalar box-integral decomposition,
\begin{align}
	\cal{M}_{n;1} = \deltfour{p} \sum_{\textrm{partitions}} 
		\left( 
		\cal{D}^{4m} I^{4m} + \cal{D}^{3m} I^{3m} + 
		\cal{D}^{2mh} I^{2mh} + \cal{D}^{2me} I^{2me} +
		\cal{D}^{1m} I^{1m}
		 \right).
\end{align}
The box integral coefficients of the scalar integrals are quadruple unitarity cuts, calculated by sewing four tree-level SUGRA amplitudes at each corner of the box,
\begin{align}
	\cal{D} = 
		\int \prod_{j=1}^4  d^8 \eta_{l_j} 
	\cal{M}(l_1,\{K_1\},-l_2) \cal{M}(l_2,\{K_2\},-l_3) 
	\cal{M}(l_3,\{K_3\},-l_4) \cal{M}(l_4,\{K_4\},-l_1).
\end{align}
The sum over partitions instructs us to include all partitions of the external particles into four subsets $K_i$, consistent with each of the different box functions in the box-integral decomposition.

Because gravity amplitudes do not possess a color-ordered structure, assigning the external particles to the corners of a box diagram requires a large number of distinct partitions.  All possible partitions of the external particles into four subsets, one for each on-shell tree amplitude in a quadruple cut, must be included to achieve the complete Bose symmetry of SUGRA amplitudes.  Furthermore, the on-shell tree amplitudes appearing as factors in the box coefficient include all permutations of the participating particles, the external legs $K_i$ and the virtual particles $l_i$ and $l_{i+1}$.  

In the context of on-shell recursion relations at tree level, this issue is confronted in Ref.~\cite{drummondsugra} by introducing ordered gravity subamplitudes $M(1,\ldots,n)$ related in \refeqn{ordered} to the physical tree amplitude $\cal{M}_n$ by adding contributions from permutations among the labels $\{2,\ldots,n-1\}$. 
 At tree level, the legs $1$ and $n$ are singled out for the complex shifts of momenta and Grassmann variables which yield on-shell recursion.  The all-multiplicity tree amplitudes in SUGRA are obtained by sewing pairs of ordered tree amplitudes and then carrying out a permutation sum over the labels of the $(n-2)$ unshifted external legs.

Instead of carrying out the permutation sum over external particles before sewing the tree amplitudes,  we prove that ordered tree-level subamplitudes may be sewn together to produce ordered one-loop quadruple cut coefficients, $D(1,\ldots,n)$.  The cut coefficients are given by applying the unitarity method for SUGRA at one-loop, that is, 
\begin{align}
	D(1,\ldots,n) \equiv 
		\int \prod_{j=1}^4  d^8 \eta_{l_j} 
	&
	M(l_1,r,\ldots,s-1,-l_2) M(l_2,s,\ldots,t-1,-l_3) 
\\ \nn \times &
	M(l_3,t,\ldots,u-1,-l_4) M(l_4,u,\ldots,r-1,-l_1), 
\end{align}
where the labels for external particles are cyclically ordered, satisfying $r<s<t<u$ modulo $n$.  The physical one-loop box functions are constructed from the ordered one-loop subamplitudes by including all permutations of the external particle labels, 
\begin{align} \label{orderedbox}
	\sum_{\textrm{partitions}} \cal{D} I
	= 
	\sum_{\sigma(1,\ldots,n)}
	\sum_{r<s<t<u}  D(1,\ldots,n)
	I(K_1,K_2,K_3,K_4). 
\end{align}
The external particles are partitioned into the sets 
$K_1=\{r,\ldots,s-1\}$, $K_2=\{s,\ldots,t-1\}$, 
$K_3=\{t,\ldots,u-1\}$, and $K_4=\{u,\ldots,r-1\}$, and the partitions are understood to be consistent with the particular box integral. 

The key ingredient in the proof is the Bose symmetry of on-shell gravity amplitudes.  Consider sewing the ordered tree-level subamplitudes and identifying the unpermuted legs $1$ and $n$ with the loop legs $l_i$ and $l_{i+1}$.  Then a permutation sum over all the external particles' labels converts each orded tree-level subamplitude into a physical amplitude.  In the proof we will focus on an arbitrary box function, 
$\sum_{\textrm{perms.}}\cal{D}_{n} I(K_1,K_2,K_3,K_4)$, but the analysis is identical for each of the box diagrams which constitute a complete one-loop amplitude.  The only caveat is that the partitioning of external legs onto corners of the box is not arbitrary but must be done in accordance with the particular quadruple-cut diagram under consideration, whether it is a four-mass box or otherwise.

The unitarity method produces a box coefficient by sewing four on-shell  tree amplitudes, as indicated in Fig.~\ref{boxdiagram}.  All the possible partitions of external legs for each corner of the box diagram are included to produce a physical amplitude, 
$\bigcup_{i=1}^4 K_i = \{1,\ldots,n\}$.  Since we are examining one box-integral coefficient, the partitions of external legs are implicitly consistent with its companion box integral.  We begin the proof by writing the box function as the product of sewn tree amplitudes and a scalar box integral, summing over all the appropriate partitions $K_i$ of external legs,
\begin{align}
	\sum_{\textrm{{\tiny parts.}}} \cal{D} I &= 
		\sum_{K_i} \int \prod_{j=1}^4  d^8 \eta_{l_j} 
	\cal{M}(l_j,\{K_j\},-l_{j+1})  \,
	I(K_1,K_2,K_3,K_4)
\\ \nn &=
	\frac{1}{n!} \sum_{\sigma(1,\ldots,n)} 
	\sum_{K_i} \int \prod_{j=1}^4  d^8 \eta_{l_j} 
	\cal{M}(l_j,\{K_j\},-l_{j+1})  \,
	I(K_1,K_2,K_3,K_4)
\\ \nn &=
	\frac{1}{n!} \sum_{\sigma(1,\ldots,n)} 
	\sum_{1<s<t<u\leq n} 
	\binom{n}{s-1} \binom{n-s+1}{t-s} \binom{n-t+1}{u-t} 
\\ \nn & \qquad \times
	\int \prod_{j=1}^4  d^8 \eta_{l_j} 
	\cal{M}(l_1,1,\ldots,s-1,-l_2) 
	\cal{M}(l_2,s,\ldots,t-1,-l_3) 
\\ \nn & \qquad  \times
	\cal{M}(l_3,t,\ldots,u-1,-l_4) 
	\cal{M}(l_4,u,\ldots,n,-l_1) \,
	I(K_i)
\\ \nn &=
	\sum_{\sigma(1,\ldots,n)} \sum_{1<s<t<u\leq n}
	\int \prod_{j=1}^4  d^8 \eta_{l_j} 
	M(l_1,1,\ldots,s-1,-l_2) M(l_2,s,\ldots,t-1,-l_3) 
\\ \nn & \qquad \times
	M(l_3,t,\ldots,u-1,-l_4) M(l_4,u,\ldots,n,-l_1) \,
	I(K_i)
\\ \nn &=
	\sum_{\sigma(1,\ldots,n)}
	\sum_{1<s<t<u\leq n} 
	D(1,\ldots,n) \, I(K_i).
\end{align}
In the second line, the complete Bose symmetry of each box function in the on-shell, one-loop amplitude is used to introduce a redundant permutation sum, 
$\sum_{\sigma(1,\ldots,n)} \cal{M}_{n;1} = n! \, \cal{M}_{n;1}$.  

Inside the permutation sum, the precise labels which are assigned to each corner by the partitioning into $K_i$ are irrelevant.  Because of the permutation sum, the only distinguishing feature of the different partitions is the number of external particles assigned to the corners.  Hence, as indicated in the third line, the sum over partitions $K_i$ inside the permutation sum is equivalent to choosing the convenient partition of external labels 
$K_1 = \{1,\ldots,s-1\},\, K_2 = \{s,\ldots,t-1\},\, 
K_3 = \{t,\ldots,u-1\},\, K_4 = \{u,\ldots,n\}$ and summing over the number of external legs which appear at each corner.  The binomial coefficients count the number of ways each distinct partition of external legs occurs, choosing $s-1$ of the $n$ particles for $K_1$ for example.  

Inside the permutation sum, all the different orderings of the external legs which lie at a given corner for fixed $s$, $t$, and $u$ yield an identical tree amplitude.  Due to the Bose symmetry of each on-shell tree amplitude in the quadruple cut, the tree amplitude $\cal{M}(l_4,1,\ldots,s-1,-l_1)$, for example, and the $(s-1)!$ permutations of its external particles are all equal.  Then we can replace each of the equivalent tree amplitudes $\cal{M}$ with an ordered subamplitude and choose numeric ordering for the external labels at each corner.  This amounts to the replacement  
$\cal{M}(l_1,1,\ldots,s-1,-l_2) 
\rightarrow (s-1)!\, M(l_1,1,\ldots,s-1,-l_2)$ inside the permutation sum, and likewise for the remaining corners of the box.  Finally, cancelling the numeric factors yields the fourth line and completes the proof that sewing ordered subamplitudes yields an ordered quadruple cut from which the complete Bose symmetric amplitude can be recovered.

The scalar box integrals of \refeqn{scalarintegral} are invariant to permutations among the elements of each individual $K_i$.  Then collecting terms in the permutation sum, \refeqn{orderedbox}, leads to a sum of ordered boxes which give the coefficient for a particular  scalar box integral.   The ordered boxes which contribute to the coefficient of a box integral differ only by relabelings of the external legs at each separate corner.
As an example,  the coefficient of the easy two-mass box integral 
$I(1,\{2,3\},4,\{5,6\})$ is the sum of the ordered boxes 
$D(1,2,3,4,5,6)$, $D(1,3,2,4,5,6)$, $D(1,2,3,4,6,5)$, and 
$D(1,3,2,4,6,5)$.

\subsection{MHV box coefficients for SUGRA}
A single two-mass easy box coefficient determines the one-loop MHV superamplitude in SUGRA, just as in SYM.  The ordered box diagram with a pair of diagonally-opposite $\mhvbarthree$ vertices yields 
\begin{align} 
&	D_{r,r+1,s,s+1}(\mhvt) =\int\prod_{i=1}^4 d^8\eta_{l_i}
\\ \nn 
& \qquad \times
	\mhvbarsugra{l_1}{r}{l_2}
	\mhvsugra{l_2}{l_3}{r+1}{s-1}
\\ \nn 
& \qquad \times
	\mhvbarsugra{l_3}{s}{l_4}
	\mhvsugra{l_4}{l_1}{s+1}{r-1}
\\  \nn
& \qquad \times
	G^{\mhvt}(l_2,r+1,\ldots,s-1,-l_3)
	G^{\mhvt}(l_4,s+1,\ldots,r-1,-l_1).
\end{align}
Since the SUGRA factors $G^{\mhvt}$ contain no $\eta$'s and are thus untouched by the Grassmann integration, the Grassmann integral simply yields the ``square" of the SYM result from Ref.~\cite{drummondloop}, 
\begin{align} 
	D_{r,r+1,s,s+1}(\mhvt) &= 
	\frac{\deltsixteen{q} }{\prod_1^n \spa{i}{i+1}^2 }
	\times
	\Delta_{r,r+1,s,s+1}^2 
\\ \nn
& \quad \times
	G^{\mhvt}(l_2,r+1,\ldots,s)
	G^{\mhvt}(l_4,s+1,\ldots,r).
\end{align}
Since $G^{\mhvt}(1,\ldots,n)$ depends on $n$ only through the spinor $\ang{n}$ and has zero phase weight in $\ang{n}$, we have used the kinematic constraints at the $\mhvbarthree$ vertices to replace $-l_3\rightarrow s$ and $-l_1\rightarrow 1$ without gaining any additional factors.

This one-loop MHV superamplitude is valid for five or more external particles.  A degeneracy of the four-point box functions doubles the result for the box coefficient.  Thus, a factor of two is required to match the result at four point from Ref.~\cite{berngraviton}.

\subsection{NMHV box coefficients for SUGRA}
First consider the diagram for an ordered three-mass box coefficient,
\begin{align} \nn
&	D_{r,r+1,s,t}^{3m}(\nmhv) =\int\prod_{i=1}^4 d^8\eta_{l_i}
\\ \nn
& \qquad \times
	\mhvbarsugra{l_1}{r}{l_2}
	\mhvsugra{l_2}{l_3}{r+1}{s-1}
\\ \nn 
& \qquad \times
	\mhvsugra{l_3}{l_4}{s}{t-1}
	\mhvsugra{l_4}{l_1}{t}{r-1}
\\ \nn 
& \qquad \times
	G^{\mhvt}(l_2,r+1,\ldots,s-1,-l_3)
	G^{\mhvt}(l_3,s,\ldots,t-1,-l_4)
\\ 
& \qquad \times
	G^{\mhvt}(l_4,t,\ldots,r-1,-l_1).
\end{align}
Again we can ``square" the SYM result and carry along the $G^{\mhvt}$'s, which are untouched by the Grassmann integrals, to obtain
\begin{align} \nn
&	D_{r,r+1,s,t}^{3m}(\nmhv) =  
	\frac{\deltsixteen{q} }{\prod_1^n \spa{i}{i+1}^2 }
	\times
	(\Delta_{r,r+1,s,t} R_{r;st})^2
\\ \nn 
& \qquad \times
	G^{\mhvt}(l_2,r+1,\ldots,s-1,-l_3)
	G^{\mhvt}(l_3,s,\ldots,t-1,-l_4) 
\\ 
& \qquad \times
	G^{\mhvt}(l_4,t,\ldots,r-1,r).
\end{align}
We have used the $\mhvbarthree$ kinematic constraint to replace $-l_1\rightarrow r$ in $G^{\mhvt}(-l_4,t,\ldots,r-1,l_1)$.  This result contains the square of the superconformal invariant $R_{r;st}$, where it is understood that the ``square" of $\deltfour{x}$ is $\delteight{x}$.

The hard two-mass box coefficients are degenerate three-mass coefficients.  The two-mass hard diagrams with massless corners $K_1$ and $K_4$ are determined by restricting the number of external legs which attach to the tree amplitudes in the three-mass diagram, 
\begin{align}
	D_{r,r+1,s,r-1}^{2mh}(\nmhv) = D_{r,r+1,s,r-1}^{3m}(\nmhv)
	+ D_{r-1,r,r+1,s}^{3m}(\nmhv).
\end{align}

The final ordered diagrams required for the NMHV SUGRA amplitude at one loop determine the two-mass easy coefficient, containing a single tree-level NMHV superamplitude.  There is no contribution to the two-mass easy coefficient obtained from restricting the three-mass box with $K_3 = \{s\}$.  As mentioned in Ref.~\cite{drummondloop}, this limit vanishes due to the kinematic constraints of three vertices.  Then we have for the two-mass easy box coefficients with the massless corners $K_1$ and $K_4$,
\begin{align} \nn
&	D_{r,r+1,s,s+1}^{2me}(\nmhv) =\int\prod_{i=1}^4 d^8\eta_{l_i}
\\ \nn
& \qquad \times
	\mhvbarsugra{l_1}{r}{l_2}
	\mhvsugra{l_2}{l_3}{r+1}{s-1}
\\ \nn 
& \qquad \times
	\mhvbarsugra{l_3}{s}{l_4}
	\mhvsugra{l_4}{l_1}{s+1}{r-1}
\\ 
& \qquad \times
	G^{\mhvt}(l_4,s+1,\ldots,-l_1)
	\times
	\sum_{a,b} R_{l_3;ab}^2 G_{l_3;ab}^{\nmhv} 
	+ (r\leftrightarrow s).
\end{align}
The summation variables $a$ and $b$ take values from the cluster $K_2$, so that $a\geq r+1$ and $a+2\leq b\leq s-1$.  The dual superconformal invariant $R_{l_3;ab}$ does not depend on $\eta_{l_2}$ or $\eta_{l_3}$, so the Grassmann integrations leaves $R_{l_3;ab}^2 G_{l_3;ab}^{\nmhv}$ untouched.  These factors are carried along with the same Grassmann integrals that appear in the MHV-amplitude unitarity cut, so from that result we have 
\begin{align} \nn
	D_{r,r+1,s,s+1}^{2me}(\nmhv) &= 
	\frac{\deltsixteen{q} }{\prod_1^n \spa{i}{i+1}^2 }
	\Delta_{r,r+1,s,s+1}^2
	G^{\mhvt}(l_4,s+1,\ldots,r)
\\ 
& \qquad \times
	\sum_{r+1\geq a,b<s} R_{s;ab}^2 G_{s;ab}^{\nmhv} 
	+ (r\leftrightarrow s).
\end{align}
Here we have used the kinematic constraints of the $\mhvbarthree$ vertices to replace $l_3\rightarrow s$ in $R_{l_3;ab}$ and $G_{l_3;ab}^{\nmhv}$, which have zero phase weight in $\ang{l_3}$, and similarly for the $G^{\mhvt}$ factor.  

The final NMHV box coefficients are the the one-mass boxes.  Taking the massive corner to be $K_4$, the one-mass coefficients are 
\begin{align}
	D_{r-2,r-1,r,r+1}^{1m}(\nmhv) = 
	  D_{r-1,r,r+1,r-2}^{3m}(\nmhv) 
	+ D_{r,r+1,r-2,r-1}^{2me}(\nmhv).
\end{align}

\subsection{NNMHV box coefficients for SUGRA}
With the NNMHV box coefficients for SYM calculated in Section \ref{subsection:nnmhvsym}, the SUGRA box functions are nearly completely determined.  Only the proper ``squaring" and insertion of the gravity $G$-factors remains to complete the SUGRA box coefficients.  We begin with the four-mass ordered box coefficient, 
\begin{align} 
&	D_{r,s,t,u}^{4m}(\nnmhv) 
	= \prod_{j=1}^{4} \int d^8\eta_{l_j} 
\\ \nn
& \qquad \times
	\mhvsugra{l_1}{l_2}{r}{s-1}
	\times
	G^{\mhvt}(l_1,r,\ldots,s-1,-l_2)
\\ \nn
& \qquad \times
	\mhvsugra{l_2}{l_3}{s}{t-1}
	\times
	G^{\mhvt}(l_2,s,\ldots,t-1,-l_3)
\\ \nn
& \qquad \times
	\mhvsugra{l_3}{l_4}{t}{u-1}
	\times
	G^{\mhvt}(l_3,t,\ldots,u-1,-l_4) 
\\ \nn
& \qquad \times
	\mhvsugra{l_4}{l_1}{u}{r-1}
	\times
	G^{\mhvt}(l_4,u,\ldots,r-1,-l_1).
\end{align}
The $G^{\mhvt}$ factors contain no Grassmann variables and factor out of the integral, leaving us with the integrand of the SYM four-mass box ``squared,"
\begin{align} \nn
	D_{r,s,t,u}^{4m}(\nnmhv) = &
	\left(	
	\frac{x_{rt}^2 x_{su}^2 }{x_{rs}^2 x_{tu}^2 }
	\Delta_{r,s,t,u}
	R_{l_3;tsu} R_{l_4;urt}
	\right)^2 
	\times \frac{\deltsixteen{q} }{\prod_1^n \spa{i}{i+1}^2 } 
\\ \nn &
	\times
	G^{\mhvt}(l_1,r,\ldots,s-1,-l_2)
	G^{\mhvt}(l_2,s,\ldots,t-1,-l_3)
\\  &
	\times
	G^{\mhvt}(l_3,t,\ldots,u-1,-l_4) 
	G^{\mhvt}(l_4,u,\ldots,r-1,-l_1).
\end{align}

Referring to Fig.~\ref{boxdiagramnnmhv} in order to calculate the ordered box coefficient $D_{r,r+1,s,t}^{I\!I\!I}(\nnmhv)$, we have  
\begin{align}
	D_{r,r+1,s,t}^{I\!I\!I}(\nnmhv) = 
			D_{r,r+1,s,t}^{I\!I\!I(A)}(\nnmhv)
			+ D_{r,r+1,s,t}^{I\!I\!I(C)}(\nnmhv) 
			+ D_{r,r+1,s,t}^{I\!I\!I(B)}(\nnmhv),
\end{align}
where, for example, the diagram $C$ yields  
\begin{align} \nn
&	D_{r,r+1,s,t}^{I\!I\!I(C)}(\nnmhv) = 
	\prod_{j=1}^{4} \int d^8\eta_{l_j} 
	\mhvbarsugra{l_2}{l_1}{r}
\\  \nn
& \qquad \times
	\mhvsugra{l_2}{l_3}{r+1}{s-1} 
	\times
	G^{\mhvt}(l_2,r+1,\ldots,s-1,-l_3) 
\\  \nn
& \qquad \times
	\mhvsugra{l_3}{l_4}{s}{t-1}
	\times
	G^{\mhvt}(l_3,s,\ldots,t-1,-l_4) 
\\ 
& \qquad \times
	\mhvsugra{l_4}{l_1}{t}{r-1}
	\times
	\sum_{t\leq a,b<r} R_{l_1;ab}^2 G^{\nmhv}_{l_1;ab} .
\end{align}
The indices $a$ and $b$ in the NMHV factor satisfy 
$a\geq t$ and $a+2\leq b \leq r-1$.  The calculation for diagrams $I\!I\!I(A-B)$ is very similar, and altogether we have 
\begin{align} \nn
	D_{r,r+1,s,t}^{I\!I\!I(A)}(\nnmhv)
	=& 
	\left( 
		\Delta_{r,r+1,s,t} R_{r;st}
	\right)^2 
	\sum_{r+1\leq a,b<s} R_{l_3;ab}^2 G^{\nmhv}_{l_4;ab} 
	\times
	\frac{\deltsixteen{q}}{\prod_1^n \spa{i}{i+1}^2}
\\ 
&  \times 
	G^{\mhvt}(l_3,s,\ldots,t-1,-l_4) 
	G^{\mhvt}(l_4,t,\ldots,r-1,r), 
\\ \nn
	D_{r,r+1,s,t}^{I\!I\!I(B)}(\nnmhv)
	=& 
	\left( 
		\Delta_{r,r+1,s,t} R_{r;st}
	\right)^2 
	\sum_{s\leq c,d<t} R_{l_4;cd}^2 G^{\nmhv}_{l_4;cd} 
	\times
	\frac{\deltsixteen{q}}{\prod_1^n \spa{i}{i+1}^2}
\\ 
&  \times 
	G^{\mhvt}(l_2,r+1,\ldots,s-1,-l_3) 
	G^{\mhvt}(l_4,t,\ldots,r-1,r), 
\\ \nn
	D_{r,r+1,s,t}^{I\!I\!I(C)}(\nnmhv)
	=& 
	\left( 
		\Delta_{r,r+1,s,t} R_{r;st}
	\right)^2 
	\sum_{t\leq a,b<r} R_{r;ab}^2 G^{\nmhv}_{r;ab} 
	\times
	\frac{\deltsixteen{q}}{\prod_1^n \spa{i}{i+1}^2}
\\ 
&  \times 
	G^{\mhvt}(l_2,r+1,\ldots,s-1,-l_3) 
	G^{\mhvt}(l_3,s,\ldots,t-1,-l_4), 
\end{align}
The kinematic constraint 
$\ang{l_1}=-\frac{\spb{r}{l_2}}{\spb{l_1}{l_2}} \ang{r}$ 
from the $\mhvbarthree$ vertex has been applied to write 
$R_{l_1;uv}=R_{r;uv}$, $G^{\nmhv}_{l_1;wz}=G^{\nmhv}_{r;wz}$, and 
$G^{\mhvt}(l_4,t,\ldots,r-1,-l_1) = G^{\mhvt}(l_4,t,\ldots,r-1,r)$ without introducing any additional factors.

The final set of ordered box coefficients for the NNMHV SUGRA amplitude are calculated from the quadruple cuts with a pair of $\mhvbarthree$ vertices.   Thus the two-$\mhvbarthree$ contributions are 
\begin{align}
D_{r,r+1,s,s+1}^{I\!I}(\nnmhv) = 
		  D_{r,r+1,s,s+1}^{I\!I(A)}(\nnmhv) 
		+ D_{r,r+1,s,s+1}^{I\!I(B)}(\nnmhv).
\end{align}
For the first diagram we have 
\begin{align} \nn
&	D_{r,r+1,s,s+1}^{I\!I(A)}(\nnmhv) = 
	\prod_{j=1}^{4} \int d^8\eta_{l_j} 
	\mhvbarsugra{l_2}{l_1}{r}
\\ \nn
& \qquad \times
	\mhvsugra{l_2}{l_3}{r+1}{s-1} 
	\times \sum_{a,b} R_{l_3;ab}^2
	G^{\nmhv}_{l_3;ab}
\\ \nn
& \qquad \times
	\mhvbarsugra{l_4}{l_3}{s}
\\ 
& \qquad \times
	\mhvsugra{l_4}{l_1}{t}{r-1} 
	\times \sum_{c,d} R_{l_1;cd}^2 
	G^{\nmhv}_{l_1;cd},
\end{align}
where the indices $a$ and $b$ in the first NMHV factor satisfy 
$a\geq r+1$ and $a+2\leq b \leq s-1$, and in the second factor we have 
$c\geq s+1$ and $c+2\leq d \leq r-1$.  Factoring the Grassmann-variable independent factors out of the integral leaves us with the MHV box coefficient, 
\begin{align} 
	D_{r,r+1,s,s+1}^{I\!I(A)}(\nnmhv) &=
	\Delta_{r,r+1,s,s+1}^2
\times
	\frac{\deltsixteen{q}}{\prod_1^n \spa{i}{i+1}^2}
\\ \nn
& \qquad \times
	\sum_{r+1\leq a,b<s} R_{s;ab}^2 G^{\nmhv}_{s;ab} 
	\times 
	\sum_{s+1\leq c,d<r} R_{r,cd}^2 G^{\nmhv}_{r,cd}.
\end{align}
The kinematic constraints at the $\mhvbarthree$ vertices have been applied to replace $l_3\rightarrow s$ and $l_1\rightarrow r$ in the NMHV tree amplitude factors.

The remaining diagrams, which contains a NNMHV SUGRA tree amplitude, are 
\begin{align} \nn
&	D_{r,r+1,s,s+1}^{I\!I(B)}(\nnmhv) = 
	\prod_{j=1}^{4} \int d^8\eta_{l_j}   
\\ \nn
& \qquad \times
	\mhvbarsugra{l_2}{l_1}{r}
	\mhvsugra{l_2}{l_3}{r+1}{s-1} 
\\ \nn
& \qquad \times
	\sum_{r+1\leq a,b< s} R_{l_3;ab}^2 
	\left[
	\sum_{a\leq c,d<b}(R_{l_3;ab;cd}^{ba})^2
	H_{l_3;ab;cd}^{(1)}+
	\sum_{b\leq c,d<n}(R_{l_3;cd}^{ab})^2
	H_{l_3;ab;cd}^{(2)}
	\right]
\\ \nn 
& \qquad \times
	\mhvbarsugra{l_4}{l_3}{s}
	\mhvsugra{l_4}{l_1}{t}{r-1} 
\\ 
	& \qquad \quad + (r\leftrightarrow s).
\end{align}
Only the MHV box coefficient remains after pulling the SUGRA NNMHV tree factor out of the integrand, leaving the expression 
\begin{align} \nn
&	D_{r,r+1,s,s+1}^{I\!I(B)}(\nnmhv) = 
	\Delta_{r,r+1,s,s+1}^2 
	\frac{\deltsixteen{q}}{\prod_1^n \spa{i}{i+1}^2}
\\ \nn
& \qquad \times
	\sum_{r+1\leq a,b< s} R_{s;ab}^2 
	\left[
	\sum_{a\leq c,d<b}(R_{s;ab;cd}^{ba})^2
	H_{s;ab;cd}^{(1)}+
	\sum_{b\leq c,d<n}(R_{s;cd}^{ab})^2
	H_{s;ab;cd}^{(2)}
	\right] 
\\ & \qquad \quad
	+(r\leftrightarrow s).
\end{align}
The kinematic constraint for the $\mhvbarthree$ vertex has allowed us to replace $\ang{l_3}\rightarrow s$.

The complete set of ordered box coefficients for NNMHV amplitudes in SUGRA are determined just as in the SYM case.  We have
\begin{align} \nn
	D_{r,r+1,s,t}^{3m} &= 
	D_{r,r+1,s,t}^{4m} + 
	D_{r,r+1,s,t}^{I\!I\!I},
\\ \nn
	D_{r,r+1,s,r-1}^{2mh} &=
	D_{r,r+1,s,r-1}^{I\!I\!I(A)} + 
	D_{r,r+1,s,r-1}^{I\!I\!I(B)} +
	D_{r-1,r,r+1,s}^{I\!I\!I(B)} + 
	D_{r-1,r,r+1,s}^{I\!I\!I(C)},
\\ \nn
	D_{r,r+1,s,s+1}^{2me} &=
	D_{r,r+1,s,s+1}^{4m} + 
	D_{r,r+1,s,s+1}^{I\!I},
\\
	D_{r-2,r-1,r,r+1}^{1m} &= 
	D_{r-1,r,r+1,r-2}^{I\!I\!I(B)} + 
	D_{r,r+1,r-2,r-1}^{I\!I(B)}.,
\end{align}
in addition to the four-mass box coefficient $D^{4m}$.

\section{Extracting gluon and graviton scattering amplitudes}
In order to use our one-loop superamplitudes to generate gluon and graviton scattering amplitudes,  the Grassmann-valued operator 
\begin{align}
	\int d\eta_i^1 \cdots d\eta_i^\cal{N}
	= \int d^\cal{N} \eta_i
\end{align}
is applied to the superamplitude.  As described above 
\refeqn{selectgluons} in the context of SYM, this operator extracts the contribution of a negative helicity gluon (for $\cal{N}=4)$ or graviton (for $\cal{N}=8$) to the scattering amplitude.  In order to compare our NNMHV results, for example, with the literature we must perform integrals on the box coefficients such as 
\begin{align}
	\int d^\cal{N} \eta_a d^\cal{N} \eta_b d^\cal{N} \eta_c 
		d^\cal{N} \eta_d \,
	\cal{C}(\textrm{NNMHV}).
\end{align}
The result of this integral is a box coefficient for a gluon or graviton scattering amplitude where the legs $a$, $b$, $c$, and $d$ have negative helicity and all the rest are positive.

Here we provide a formula which makes these integrations straightforward.  The general Grassmann dependence of a NNMHV box coefficient is in the product of the overall supercharge delta function with a pair of $\cal{N}$-component delta functions.  We are concerned only with the appearance of $\eta_i$ for $i=a,b,c,d$, and we have schematically
\begin{align} \nn
\cal{C}(\textrm{NNMHV}) &= X \times 
	\delta^{2\cal{N}}
		(\susyq{a}+\susyq{b}+\susyq{c}+\susyq{d}+\ldots)
\\ & \quad \times
	\delta^{\cal{N}}(A\eta_a +B\eta_b+\ldots)
	\delta^{\cal{N}}(C\eta_a +D\eta_b+\ldots), 
\end{align}
where $X$ is an overall bosonic factor and the $(\ldots)$ indicate Grassmann variables other than the ones of interest.  The overall supercharge delta function has been used to write the latter pair of delta functions so that they each depend on only two of the four Grassmann variables of interest.  Then the identity  \refeqn{deltaeightfactor} is applied to carry out the Grassmann integrals, and we find
\begin{align}
	\int d^\cal{N} \eta_a d^\cal{N} \eta_b d^\cal{N} \eta_c 
		d^\cal{N} \eta_d \,
	\cal{C}(\textrm{NNMHV}) =
	\spa{c}{d}^\cal{N} (AD-CB)^\cal{N} X.
\end{align}
The detailed form of the spinor products $A$, $B$, $C$, and $D$ depends on the particular box coefficient under consideration.

\section{Conclusion}
We have applied generalized unitarity to calculate the all-multiplicity, NNMHV contributions to one-loop scattering amplitudes in maximally supersymmetric Yang-Mills and Supergravity.  Our results for the NNMHV box coefficients in SYM are expressed in a manifestly dual superconformal form \cite{drummondloop}, reflecting the proposed duality between dual Wilson loops and SYM scattering amplitudes at weak coupling.  The interesting effects of collinear and infrared processes on conformal symmetry \cite{exacting} requires further study.  Does the combination of superconformal and conjectured dual superconformal symmetries \cite{yangian} fix the form of scattering amplitudes in $\cal{N}=4$ Yang-Mills?

We also calculated the $n$-point MHV and NMHV one-loop amplitudes in SUGRA.  The requisite tree amplitudes which are sewn together for the coefficients of quadruple cuts were calculated in \cits{drummondtrees,drummondsugra}.  The amplitudes we present are generating functions for the scattering of any of the particles in the supermultiplet appearing as external states.  The use of an on-shell superspace formalism described in Ref.~\cite{drummondsym} allows scattering amplitudes with external gluon or graviton states to be easily extracted with Grassmann-valued operators.  

In this paper we proved that the ordered gravity subamplitudes introduced in Ref.~\cite{drummondsugra} may be sewn together to produce  ordered box coefficients from which the complete, physical box coefficients are obtained by permuting all the external legs.  The ordered subamplitudes are not physical quantities, but they do yield a more efficient means of calculating SUGRA amplitudes.  Instead of permuting over all the legs for each tree amplitude in a quadruple cut, the ordered trees may be sewn directly and then permuted over only the external legs at the end.  This allows the box coefficients to be represented concisely, for arbitrary multiplicity scattering processes.

Here we describe the checks we have performed on our results.
The results for SYM box coefficients have been numerically checked  against the box coefficients for the amplitude 
$\cal{A}_{7;1}(1^-,2^-,3^-,4^+,5^+,6^+,7^+)$ presented in Ref.~\cite{bernsevengluon}.  The complex conjugate of this seven-gluon amplitude can be considered a NNMHV amplitude and allows a non-trivial verification of our box coefficients.  We note that the coefficients presented in Ref.~\cite{bernsevengluon} multiply box functions instead of the box integrals alone, where the box functions are scalar box integrals multiplied by the appropriate $\Delta_{r,s,t,u}$.  Thus in comparing our results we find relations between our  coefficients such as
\begin{align}
	c_{267}^\ast = \int 
		d^4\eta_4\, d^4\eta_5\, d^4\eta_6\, d^4\eta_7 
			\frac{\cal{C}_{3,4,5,1}^{I\!I\!I(B)} }
			{\Delta_{3,4,5,1} }.
\end{align}

For the SUGRA box coefficients we have numerically verified that the MHV one-loop amplitudes for the scattering of four, five, and six gravitons matches the results in Ref.~\cite{berngraviton} up to overall normalization of the amplitudes.  The complex conjugate of the five-point MHV coefficients match our NMHV result, and likewise the six-point MHV amplitude, upon conjugation, agrees with our NNMHV box coefficients.  Recall that the ordered box coefficients we calculated yield physical SUGRA amplitudes by summing over permutations of the external legs.  Then, after this permutation sum is carried out, the physical amplitudes for the scattering of gravitons can be extracted by Grassmann-valued operators.  For instance, we find the relation 
\begin{align} \nn
&	\left[\spa{1}{2}^8 h(1,\{2,3\},4) h(4,\{5,6\},1) \right]^\ast
	\textrm{tr}^2[1(2+3)4(5+6)]
\\ & \qquad =
	\int 
	d^8\eta_3\, d^8\eta_4\, d^8\eta_5\, d^8\eta_6
	\sum_{\sigma(23),\sigma(56)} D_{1,2,4,5}^{4m}.
\end{align}
The MHV amplitudes we have checked against have few external legs and thus do not require all the box functions which appear in the all-multiplicity NNMHV boxes.  Nevertheless, they provide an affirmation of our method for producing physical amplitudes from ordered box coefficients.

\section*{Acknowledgments}
The author is grateful to Zvi Bern for guidance, especially in the process of checking the results presented in this paper.  The author thanks Academic Technology Services at UCLA for the computer support which made the programming and numerical analysis of amplitudes possible.

\begingroup\raggedright

\endgroup

\end{document}